\documentclass[12pt, a4paper]{article}
\usepackage{graphicx}
\usepackage{amssymb}
\usepackage{amsmath}
\usepackage{bm}
\usepackage{color}
\usepackage{theorem}
\usepackage{subcaption}
\usepackage{listings}
\usepackage{colortbl}
\usepackage{tabularx}
\usepackage{longtable}
\usepackage[utf8]{inputenc}
\usepackage[T1]{fontenc}
\usepackage{lmodern}
\usepackage[top=30truemm,bottom=30truemm,left=25truemm,right=25truemm]{geometry}

\usepackage[sort&compress,numbers, merge]{natbib}

\definecolor{Orange}{cmyk}{0,0.61,0.87,0}
\definecolor{JungleGreen}{cmyk}{0.99,0,0.52,0}
\definecolor{OliveGreen}{cmyk}{0.64,0,0.95,0.40}
\definecolor{Brown}{cmyk}{0,0.81,1,0.60}
\definecolor{RoyalBlue}{cmyk}{0.71,0.53,0,0.12}
\definecolor{Gray}{cmyk}{0,0,0,0.40}
\definecolor{LightPink}{cmyk}{0.0,0.25,0,0}
\definecolor{LLightPink}{cmyk}{0.0,0.10,0,0}
\definecolor{LightBlue}{cmyk}{0.25,0,0,0}
\definecolor{LightGray}{cmyk}{0,0,0,0.2}

\allowdisplaybreaks[1]

\usepackage[colorlinks=true, linkcolor=OliveGreen, citecolor=RoyalBlue,
urlcolor=RoyalBlue]{hyperref}

\renewcommand{\thefootnote}{\fnsymbol{footnote}}

\begin{document}

\begin{titlepage}

  \begin{flushright}
\end{flushright}

\vskip 1.35cm
\begin{center}

{\LARGE
{\bf
$R$-Symmetric Flipped SU(5) 
}
}

\vskip 1.5cm

Koichi Hamaguchi$^{a,b}$\footnote{
  E-mail address: \href{mailto:hama@hep-th.phys.s.u-tokyo.ac.jp}{\tt hama@hep-th.phys.s.u-tokyo.ac.jp}}, 
Shihwen Hor$^a$\footnote{
  E-mail address: \href{mailto:shihwen@hep-th.phys.s.u-tokyo.ac.jp}{\tt shihwen@hep-th.phys.s.u-tokyo.ac.jp}},
Natsumi Nagata$^a$\footnote{
E-mail address: \href{mailto:natsumi@hep-th.phys.s.u-tokyo.ac.jp}{\tt natsumi@hep-th.phys.s.u-tokyo.ac.jp}}

\vskip 0.8cm

{\it $^a$Department of Physics, University of Tokyo, Bunkyo-ku, Tokyo
 113--0033, Japan} \\[2pt]
 {\it ${}^b$Kavli Institute for the Physics and Mathematics of the
 Universe (Kavli IPMU), University of Tokyo, Kashiwa 277--8583, Japan} 

\date{\today}

\vskip 1.5cm

\begin{abstract}

  We construct a supersymmetric flipped SU(5) grand unified model that possesses an $R$ symmetry. This $R$ symmetry forbids dangerous non-renormalizable operators suppressed by a cut-off scale up to sufficiently large mass dimensions so that the SU(5)-breaking Higgs field develops a vacuum expectation value of the order of the unification scale along the $F$- and $D$-flat directions, with the help of the supersymmetry-breaking effect.  The mass terms of the Higgs fields are also forbidden by the $R$ symmetry, with which the doublet-triplet splitting problem is solved with the missing partner mechanism. The masses of right-handed neutrinos are generated by non-renormalizable operators, which then yield a light neutrino mass spectrum and mixing through the seesaw mechanism that are consistent with neutrino oscillation data. This model predicts one of the color-triplet Higgs multiplets to lie at an intermediate scale, and its mass is found to be constrained by proton decay experiments to be $\gtrsim 5 \times 10^{11}$~GeV. If it is $\lesssim 10^{12}$~GeV, future proton decay experiments at Hyper-Kamiokande can test our model in the $p \to \pi^0 \mu^+$ and $p \to K^0 \mu^+$ decay modes, in contrast to ordinary grand unified models where $p \to \pi^0 e^+ $ or $p \to K^+ \bar{\nu}$ is the dominant decay mode. This characteristic prediction for the proton decay branches enables us to distinguish our model from other scenarios. 

\end{abstract}

\end{center}
\end{titlepage}

\renewcommand{\thefootnote}{\arabic{footnote}}
\setcounter{footnote}{0}

\section{Introduction}
\label{sec:intro}

Quarks and leptons in the Standard Model (SM) may be unified at high energies. Since 1970's, there have been many efforts to construct a concrete model that accommodates the unification of quarks and leptons, as well as of the strong and electroweak interactions~\cite{Pati:1973uk, Pati:1974yy, Georgi:1974sy, Georgi:1974my, Fritzsch:1974nn, Gursey:1975ki}. Most of these models---dubbed as grand unified theories (GUTs)---predict that the unification is achieved at a very high energy~\cite{Georgi:1974yf}, and the large difference between the electroweak and unification scales brings about the hierarchy problem~\cite{Weinberg:1974tw, Weinberg:1975gm, Gildener:1976ih, Gildener:1976ai, Susskind:1978ms}. The electroweak scale may be stabilized against this large hierarchy if supersymmetry (SUSY) appears above the electroweak scale, which highly motivates SUSY GUTs~\cite{Dimopoulos:1981zb,Sakai:1981gr}. In particular, the minimal SUSY SU(5) GUT~\cite{Dimopoulos:1981zb,Sakai:1981gr} has been regarded as a representative, promising model of GUTs as the matter content in the minimal SUSY SM (MSSM) nicely fits SU(5) representations and, among other things, gauge coupling unification is found to be achieved with great accuracy~\cite{Dimopoulos:1981yj, Marciano:1981un, Einhorn:1981sx, Amaldi:1991cn, Langacker:1991an, Ellis:1990wk, Giunti:1991ta}.

A drawback of the minimal SU(5) is that a high level of fine-tuning is required to keep the MSSM Higgs fields from acquiring a GUT-scale mass while making their SU(5) partner fields massive. This problem, called the doublet-triplet splitting problem, is a generic problem in GUTs, and many solutions to this problem have been proposed so far. Among them, the missing partner mechanism~\cite{Masiero:1982fe, Grinstein:1982um} is most frequently discussed in SU(5) GUTs, where a {\bf 75} and a pair of ${\bf 50}$ and $\overline{\bf 50}$ representations of SU(5) are introduced in addition to the ${\bf 10}$ and $\overline{\bf 5}$ representations for the MSSM matter fields and ${\bf 5}$ and $\overline{\bf 5}$ for the MSSM Higgs fields. The SU(5) gauge symmetry is spontaneously broken by a vacuum expectation value (VEV) of the {\bf 75} field and it also gives vector-like masses to the color-triplet components in the ${\bf 5}$ and $\overline{\bf 5}$ Higgs fields together with the corresponding components in the ${\bf 50}$ and $\overline{\bf 50}$ fields, whereas the doublet components in the ${\bf 5}$ and $\overline{\bf 5}$ Higgs fields, corresponding to the MSSM Higgs fields, do not acquire masses from the VEV of the ${\bf 75}$. In order for this mechanism to work, the absence of the bilinear term for the ${\bf 5}$ and $\overline{\bf 5}$ Higgs fields is crucial, which may be attributed to an additional symmetry~\cite{Hisano:1992ne, Hisano:1994fn, Berezhiani:1996nu, Altarelli:2000fu, Antusch:2014poa, Bjorkeroth:2015ora}. In spite of the conceptual success of the missing partner mechanism, it is known that the missing partner model with the minimal matter content is incompatible with the perturbative gauge coupling unification due to the existence of large representations~\cite{Pokorski:2019ete} and thus a more elaborate model building is required to realize this mechanism in SU(5). 
 
Another economical way to realize this missing partner mechanism is found in the flipped ${\rm SU}(5) \times {\rm U}(1)$ model~\cite{Barr:1981qv, Derendinger:1983aj, Antoniadis:1987dx, Barr:1988yj}, where the right-handed charged leptons and neutrinos, and the right-handed up- and down-type quarks, are flipped with respect to the standard SU(5) assignment. In this model, the ${\rm SU}(5) \times {\rm U}(1)$ gauge symmetry is broken by the VEVs of ${\bf 10}$ and $\overline{\bf 10}$ Higgs fields, and they give masses only to the color-triplet components in the ${\bf 5}$ and $\overline{\bf 5}$ Higgs fields~\cite{Antoniadis:1987dx}. As a result, if there is no bilinear term of the ${\bf 5}$ and $\overline{\bf 5}$ Higgs fields, then the MSSM Higgs fields remain massless after the ${\rm SU}(5) \times {\rm U}(1)$ symmetry is spontaneously broken. Another feature of this model is that the ${\bf 10}$ and $\overline{\bf 10}$ Higgs fields develop VEVs along the $F$- and $D$-flat directions with the help of the effect of soft masses as well as non-renormalizable operators. Such non-renormalizable operators must be suppressed up to sufficiently large mass dimensions in order for these VEVs to be of the order of the unification scale. The suppression of these operators, as well as the absence of the bilinear term of the ${\bf 5}$ and $\overline{\bf 5}$ Higgs fields, are just assumed in the standard flipped SU(5) model~\cite{Antoniadis:1987dx}---it is, therefore, desirable to account for these key assumptions by means of certain symmetries. 

In this work, we construct a flipped SU(5) model that can resolve this issue. To that end, we require the model to have a global U(1)$_R$ symmetry which forbids all of the unwanted terms mentioned above. This $R$ symmetry is supposed to be explicitly broken only by the constant term in the superpotential so that we can obtain a viable SUSY mass spectrum as well as the vanishing cosmological constant. We then show that the U(1)$_R$ and ${\rm SU}(5) \times {\rm U}(1)$ symmetries are broken at the scales of $\simeq 3 \times 10^{17}$~GeV and $\simeq 10^{16}$~GeV, respectively, and that the $\mu$-term of the MSSM Higgs doublets is naturally suppressed thanks to the missing partner mechanism.

As it turns out, this model predicts one of the color-triplet Higgs multiplets to lie at an intermediate scale. We evaluate the current limit on the mass of this light color-triplet Higgs from proton decay experiments and discuss its implications for the future experiments at Hyper-Kamiokande. It is found that in the presence of a color-triplet Higgs with a mass $\lesssim 10^{12}$~GeV, proton decay can be discovered at Hyper-Kamiokande in the $p \to \pi^0 \mu^+$ and $p \to K^0 \mu^+$ decay modes. Note that in ordinary GUT models either $p \to \pi^0 e^+ $ or $p \to K^+ \bar{\nu}$ is the dominant decay mode. We may therefore distinguish our model from others with proton decay experiments, which highly motivates to search for the $p \to \pi^0 \mu^+$ and $p \to K^0 \mu^+$ modes in future proton decay experiments, in addition to the standard channels $p \to \pi^0 e^+ $ and $p \to K^+ \bar{\nu}$. 

This paper is organized as follows. In the subsequent section, we describe our model and discuss its symmetry breaking structure and mass spectrum. In Sec.~\ref{sec:flavor}, we study the flavor structure of this model, and show that the observed pattern of the SM quark and lepton masses and mixings can be reproduced. We then compute the proton decay rates in Sec.~\ref{sec:protondecay} and discuss the testability of our model. Finally, Sec.~\ref{sec:conclusion} is devoted to conclusion and discussion.

\section{$R$-symmetric flipped SU(5)}
\label{sec:model}

\subsection{Model}
\label{sec:lagrangian}

The model discussed in this paper is based on an ${\rm SU}(5) \times {\rm U}(1)$ gauge theory~\cite{Barr:1981qv, Derendinger:1983aj, Antoniadis:1987dx, Barr:1988yj}, where the three generations of MSSM matter fields, as well as three right-handed neutrino chiral superfields, are embedded into $\mathbf{10}\,(1)$, $\bar{\mathbf{5}}\,(-3)$ and $\mathbf{1}\,(5)$ representations, with the numbers in the parentheses indicating the U(1) charges in units of $1/\sqrt{40}$. We denote these representations by $F_i$, $\bar{f}_i$, and $\ell^c_i$, respectively, with $i=1,2,3$ the generation index. In addition to these matter fields, this model contains a pair of $\mathbf{10}\, (1)$ and $\overline{\mathbf{10}}\, (-1)$ Higgs fields, $H$ and $\bar{H}$, a pair of $\mathbf{5}\,(-2)$ and $\overline{\mathbf{5}}\, (2)$ Higgs fields, $h$ and $\bar{h}$, and a singlet field, $S$. The $H$ and $\bar{H}$ fields break the ${\rm SU}(5) \times {\rm U}(1)$ gauge group down to the SM gauge group once these fields develop VEVs. As in the MSSM, the $\mathrm{SU}(2)_L \times \mathrm{U}(1)_Y$ gauge symmetry is broken by the VEVs of the doublet Higgs fields $H_d$ and $H_u$, which reside in $h$ and $\bar{h}$, respectively. In Table~\ref{tab:model}, we summarize the field content and the charge assignments of the fields. With these charge assignments, the U(1)$_Y$ hypercharge $Y$ is given by the following linear combination of the SU(5) generator $T_{24} = \mathrm{diag}(2,2,2,-3,-3)/\sqrt{60}$ and the U(1) charge $Q_X$:\footnote{We adopt the normalization of hypercharge such that left-handed neutrinos have hypercharge $-1/2$. We also use the SU(5) normalization of hypercharge, $Q_1 \equiv \sqrt{3/5}\, Y$. The gauge coupling constants corresponding to $Y$ and $Q_1$ are denoted by $g^\prime$ and $g_1$, respectively.  } 
\begin{equation}
    Y = \frac{1}{\sqrt{15}} T_{24} + \sqrt{\frac{8}{5}} Q_X ~.
\end{equation}
The second column in Table.~\ref{tab:model} summarizes the component fields in each representation; for instance, $F_i$ and $\bar{f}_i$ contain the MSSM matter fields as\footnote{In these equations, we neglect the mixing among different generations just for simplicity. More complete expressions for the embedding of the component fields are given in Sec.~\ref{sec:flavor}. } 
\begin{equation}
  F_i = \frac{1}{\sqrt{2}} 
  \begin{pmatrix}
    0 & d^c_{i 3} & -d^c_{i2} & u^1_i & d^1_i \\
    -d^c_{i 3} & 0& d^c_{i1} & u^2_i & d^2_i \\
    d^c_{i2} & - d^c_{i1} & 0 & u^3_i & d^3_i \\
    -u^1_i & -u^2_i & -u^3_i & 0 & \nu^c_i \\
    -d^1_i & -d^2_i & -d^3_i & -\nu^c_i & 0
  \end{pmatrix}
  ~, \qquad
  \bar{f}_i = 
  \begin{pmatrix}
    u^c_{i 1} \\ u^c_{i 2} \\ u^c_{i 3} \\ e_i \\ -\nu_i 
  \end{pmatrix}
  ~.
\end{equation}
Note that the right-handed neutrino fields $\nu^c_i$ are necessary ingredients in flipped SU(5) GUT models to form complete ${\bf 10}$ representations, in contrast to the standard SU(5) GUT~\cite{Georgi:1974sy}, where right-handed neutrino fields are singlet under the SU(5) gauge group.

\begin{table}[t]
  \centering
  \begin{tabular}{l l ccc}
  \hline \hline
  Fields & Components & SU(5)\quad & U(1)\quad & U(1)$_R$\quad  \\
  \hline
  $F_i$ & $d^c_i$, $Q_i$, $\nu^c_i$ & $\mathbf{10}$ & $+1$ & $17/36$ \\ 
  $\bar{f}_i$ & $u^c_i$, $L_i$ & $\overline{\mathbf{5}}$ & $-3$ &$17/36$\\ 
  ${\ell}_i^c$ & $e^c_i$ & ${\mathbf{1}}$ & $+5$ &$17/36$\\ 
  $H$ & $d^c_H$, $Q_H$, $\nu^c_H$ & $\mathbf{10}$ & $+1$ &$1/36$\\ 
  $\bar{H}$ & ${d}^c_{\bar{H}}$, ${Q}_{\bar{H}}$, ${\nu}^c_{\bar{H}}$ & $\overline{\mathbf{10}}$ & $-1$  &$17/36$\\ 
  $h$ & $D$, $H_d$ & $\mathbf{5}$ & $-2$ &$19/18$\\ 
  $\bar{h}$ & $\bar{D}$, $H_u$ & $\overline{\mathbf{5}}$ & $+2$ &$19/18$\\ 
  $S$ & $S$ & ${\mathbf{1}}$ & $0$ &$1/9$\\ 
   \hline
  \hline
  \end{tabular}
  \caption{
   The field content and the charge assignments in our model. The U(1) charges are given in units of $1/\sqrt{40}$. 
  }
  \label{tab:model}
  \end{table}
  
In addition, we assume that this model respects a global U(1)$_R$ symmetry in the global SUSY limit. The U(1)$_R$ charge assignment is also shown in Table~\ref{tab:model}, where we normalize the U(1)$_R$ charge such that the superpotential has the U(1)$_R$ charge $+2$. As we see below, this $R$ symmetry plays an important role in suppressing unwanted terms. We can instead consider a discrete $R$ symmetry with the same matter content to suppress unwanted terms; we show an example for this case in Appendix~\ref{sec:discreter}. In either case, the $R$ symmetry is broken by the VEV of the singlet field $S$.

We note that this model contains only one singlet field, $S$, contrary to the standard flipped SU(5) model~\cite{Antoniadis:1987dx, Ellis:2017jcp, Ellis:2018moe, Ellis:2019jha, Ellis:2019opr}, where several singlet fields are introduced to generate neutrino masses via the double seesaw mechanism~\cite{Minkowski:1977sc, Yanagida:1979as, Glashow:1979nm, GellMann:1980vs, Mohapatra:1979ia}; namely, our model has a simpler matter content than the ordinary one. In the present case, the masses of the right-handed neutrinos are generated via non-renormalizable operators~\cite{Antoniadis:1988tt}, as we see below.

The superpotential terms allowed by these symmetries are 
\begin{equation}
  W = W_{\rm Yukawa} + W_{\rm DT} +W_{\rm neutrino}+ W_{HS} \dots  ~,
  \label{eq:superpotential}
\end{equation}
with 
\begin{align}
  W_{\rm Yukawa} &=\frac{1}{4} \lambda_1^{ij}\epsilon_{\alpha\beta\gamma\delta\epsilon} F_i^{\alpha \beta} F_j^{\gamma \delta} h^{\epsilon}_{} + \sqrt{2} \lambda_2^{ij} F_i^{\alpha \beta}\bar{f}_{j\alpha}\bar{h}_{\beta} + \lambda_3^{ij}\bar{f}_{i\alpha}\ell^c_j h^{\alpha} ~,\label{eq:wyukawa}\\[2pt]
  W_{\rm DT} &= \frac{\lambda_4}{4\Lambda_{\rm DT}^8} \epsilon_{\alpha\beta\gamma\delta\epsilon} S^8 H^{\alpha\beta}H^{\gamma\delta} h^{\epsilon} + \frac{1}{4} \lambda_5\epsilon^{\alpha\beta\gamma\delta\epsilon}  \bar{H}_{\alpha\beta}\bar{H}_{\gamma\delta} \bar{h}_{\epsilon} ~, 
  \label{eq:wdt}
  \\[2pt]
  W_{\rm neutrino} &=  \frac{c_{ij}}{2\Lambda_{N}^2} S (F_i^{\alpha\beta}\bar{H}_{\alpha\beta} )(F_j^{\gamma\delta}\bar{H}_{\gamma\delta})~, 
  \label{eq:wneutrino}
  \\[2pt]
  W_{HS} &= \frac{\lambda_{H}}{4\Lambda_{HS}^{5}} (H^{\alpha\beta} \bar{H}_{\alpha\beta})^4 
   + \frac{\lambda_{HS}}{18 \Lambda_{HS}^{10}} (H^{\alpha\beta} \bar{H}_{\alpha\beta})^2 S^9 + \frac{\lambda_S}{18 \Lambda_{HS}^{15}} S^{18} ~, 
   \label{eq:whs}
\end{align}
where the Greek superscripts and subscripts denote the SU(5) indices, $\epsilon_{\alpha \beta\gamma\delta\epsilon}$ is the totally antisymmetric tensor, and the dots in Eq.~\eqref{eq:superpotential} indicate higher-dimensional operators that are irrelevant to the following discussions.\footnote{Exceptions are those who cause $R$-parity violation and dimension-five proton decay operators at low energies, which we discuss in Sec.~\ref{sec:rparityviolation} and in Sec.~\ref{sec:protondecay}, respectively.} $\Lambda_{\rm DT}$, $\Lambda_N$, and $\Lambda_{HS}$ are parameters with a mass dimension one that correspond to a cut-off scale of this model. Considering that the non-renormalizable operators in Eqs.~(\ref{eq:wdt}--\ref{eq:whs}) depend on a large power of the cut-off scale and that different types of fields and interactions may be responsible for generating these operators, we have introduced different parameters for the cut-off scales of these operators, though we expect all of them to be around the reduced Planck mass scale, $M_P$.  The terms in Eq.~\eqref{eq:wyukawa} contain the MSSM Yukawa coupling terms. After the $\text{SU}(5) \times \text{U}(1)$ and U(1)$_R$ symmetries are broken, $W_{\rm DT}$ gives masses to the color-triplet Higgs fields and $W_{\rm neutrino}$ leads to a Majorana mass matrix for right-handed neutrinos, as we see in Sec.~\ref{sec:massspectrum}. $W_{HS}$ in Eq.~\eqref{eq:whs} includes all of the terms that are composed of the $H$, $\bar{H}$, and $S$ fields and allowed by the symmetries of the theory; as we see, the U(1)$_R$ symmetry highly restricts possible forms of such operators. We take $\lambda_H$ and $\lambda_S$ to be real and positive without loss of generality. Notice that the bilinear terms $H\bar{H}$ and $h\bar{h}$ are both forbidden by the U(1)$_R$ symmetry.\footnote{The terms $F_i \bar{H}$ are also prohibited by the the U(1)$_R$ symmetry. } In the standard flipped SU(5) model, the $H\bar{H}$ term is forbidden by a $\mathbb{Z}_2$ symmetry, while the $h\bar{h}$ term is generically allowed. As we see in Sec.~\ref{sec:massspectrum}, the absence of the $h\bar{h}$ term is crucial to solve the doublet-triplet splitting problem.

In the following analysis, we assume that the coefficients of the operators in Eqs.~(\ref{eq:wdt}--\ref{eq:whs}) are ${\cal O}(1)$; \textit{i.e.}, the hierarchy in the contribution of these operators is completely controlled by the symmetries of the model. For the terms in Eq.~\eqref{eq:wyukawa}, on the other hand, we allow the coefficients to be much smaller than ${\cal O}(1)$ to reproduce the SM Yukawa couplings. Although the structure of the SM Yukawa couplings may also be explained by using the U(1)$_R$ symmetry with a generation-dependent charge assignment---in a similar manner to the Froggatt--Nielsen model~\cite{Froggatt:1978nt}---we do not pursue this possibility in the present paper and defer it to future work~\cite{HHN}.

\subsection{Symmetry breaking and mass spectrum}
\label{sec:massspectrum}

Next, we study the vacuum structure of this model. In the following analysis, we assume the canonical form of the K\"{a}hler potential, $K = |F_i^{\alpha\beta}|^2 + \dots$, just for simplicity. In this case, the scalar potential of the theory is computed as 
\begin{equation}
  V = V_F + V_D +V_{\rm soft} ~,
  \label{eq:scapot}
\end{equation}
with
\begin{align}
  V_F &= \sum_{i} \, \biggl| \frac{\partial W}{\partial \phi_i} \biggr|^2 ~, \\[2pt]
  V_D &= \frac{g_5^2}{2} \bigl[\sum_{i} \phi^{*i} T^A_i \phi_i \bigr]^2 
  + \frac{g_X^2}{2} \bigl[\sum_{i} Q_{X_i} |\phi_i|^2 \bigr]^2 ~,
\end{align}
where $\phi_i$ denote the scalar components of the chiral superfields in this model, $g_5$ and $g_X$ are the gauge coupling constants of the SU(5) and U(1) gauge interactions, respectively, $T^A_{i}$ and $Q_{X_{i}}$ are the representation of the SU(5) generators and U(1)$_X$ charge of the field $\phi_i$, respectively, and $V_{\rm soft}$ denotes the contribution of soft SUSY-breaking terms to the scalar potential. 

We search for a minimum of the potential at which $\nu^c_H$, $\nu^c_{\bar{H}}$, and $S$ develop VEVs; the SM gauge group is unbroken in this case. In fact, it turns out that there is a $F$- and $D$-flat direction for these fields in the absence of SUSY-breaking effect and non-renormalizable operators. To see this, let us set all of the fields other than these three fields to be zero and $V_{\rm soft} = 0$; in this case, at renormalizable level, $V_F = 0$ and 
\begin{equation}
  V_D = \biggl(\frac{3g_5^2}{10} + \frac{g_X^2}{80}\biggr) \left[ 
  \left| \nu^c_H\right|^2  -\left|\nu^c_{\bar{H}}\right|^2 
  \right]^2 ~.
  \label{eq:vdsimp}
\end{equation} 
Therefore, the potential vanishes for $| \nu^c_H| = |\nu^c_{\bar{H}}| $ and arbitrary values of $S$. 

We now include the effect of non-renormalizable operators and soft SUSY breaking masses, with which the potential terms for the above fields are 
\begin{align}
  V_F &= 
  \biggl| \frac{\lambda_H}{\Lambda_{HS}^5} (\nu^c_H  \nu^c_{\bar{H}})^2 
   + \frac{\lambda_{HS}}{9\Lambda_{HS}^{10}} S^9 
  \biggr|^2 \left| \nu^c_H  \nu^c_{\bar{H}}\right|^2 
  \left(\left| \nu^c_H \right|^2 +\left|\nu^c_{\bar{H}}\right|^2   \right)
  \nonumber \\
  &+ \left|
  \frac{\lambda_{HS}}{2\Lambda_{HS}^{10}} (\nu^c_H  \nu^c_{\bar{H}})^2 
  + \frac{\lambda_S}{\Lambda_{HS}^{15}} S^9
  \right|^2 \left|S\right|^{16} ~,\label{eq:vfsimp} \\[3pt]
  V_{\rm soft} &= - m_H^2 \left| \nu^c_H \right|^2  - m_{\bar{H}}^2 
  \left|\nu^c_{\bar{H}}\right|^2 - m_S^2 \left|S\right|^2 ~,
\end{align}
where we assume $m_H^2$, $m_{\bar{H}}^2$, and $m_S^2$ are all positive. Let us search for a minimum of the potential $V$ in Eq.~\eqref{eq:scapot} in the flat direction of the $D$-term potential in Eq.~\eqref{eq:vdsimp}; \textit{i.e.}, $| \nu^c_H| = |\nu^c_{\bar{H}}| $. To that end, we first note that by using the U(1) gauge and U(1)$_R$ transformations, we can always take the VEV of $S$ to be real and positive, and the VEVs of $\nu^c_H$ and $\nu^c_{\bar{H}}$ to satisfy $\langle \nu^c_H\rangle = \langle \nu^c_{\bar{H}}\rangle = \Phi$. The field $\Phi$ is sometimes called flaton~\cite{Ellis:2017jcp, Ellis:2018moe, Ellis:2019jha, Ellis:2019opr}. Now recall that we have taken $\lambda_H$ and $\lambda_S$ to be real and positive. To estimate the size of the VEVs of $\Phi$ and $S$ with a simple analysis, we also take $\lambda_{HS}$ to be real and positive. In this case, we can express the potential in terms of $\Phi \equiv |\Phi|e^{i \theta}$ and $S$ as 
\begin{align}
  V&= 2 |\Phi|^6 \biggl[\frac{\lambda_H^2}{\Lambda_{HS}^{10}} |\Phi|^8 +  \frac{\lambda_{HS}^2}{81\Lambda_{HS}^{20}} S^{18} + \frac{2\lambda_H \lambda_{HS} }{9 \Lambda_{HS}^{15}  } S^9 |\Phi|^4 \cos (4\theta) \biggr] \nonumber \\[2pt] 
  &+ S^{16} \biggl[\frac{\lambda_{HS}^2}{4\Lambda_{HS}^{20}} |\Phi|^8 +  \frac{\lambda_{S}^2}{\Lambda_{HS}^{30}} S^{18} + \frac{2\lambda_{HS} \lambda_{S} }{2 \Lambda_{HS}^{25}  } S^9 |\Phi|^4 \cos (4\theta) 
  \biggr] \nonumber \\[2pt]
  &- m_\Phi^2 |\Phi|^2 - m_S^2 S^2 ~,
  \label{eq:vofphiands}
\end{align}
with $m_\Phi^2 \equiv m_H^2 + m_{\bar{H}}^2$. As we see, the potential minimum is obtained for $\cos (4\theta) = -1$, meaning that at the minimum of the potential, $\Phi^4$ is real and negative. We then solve the minimization conditions of the potential for a positive value of $S$ and a negative value of $\Phi^4$ to determine their VEVs.

\begin{figure}
  \centering
  {\includegraphics[width=0.6\textwidth]{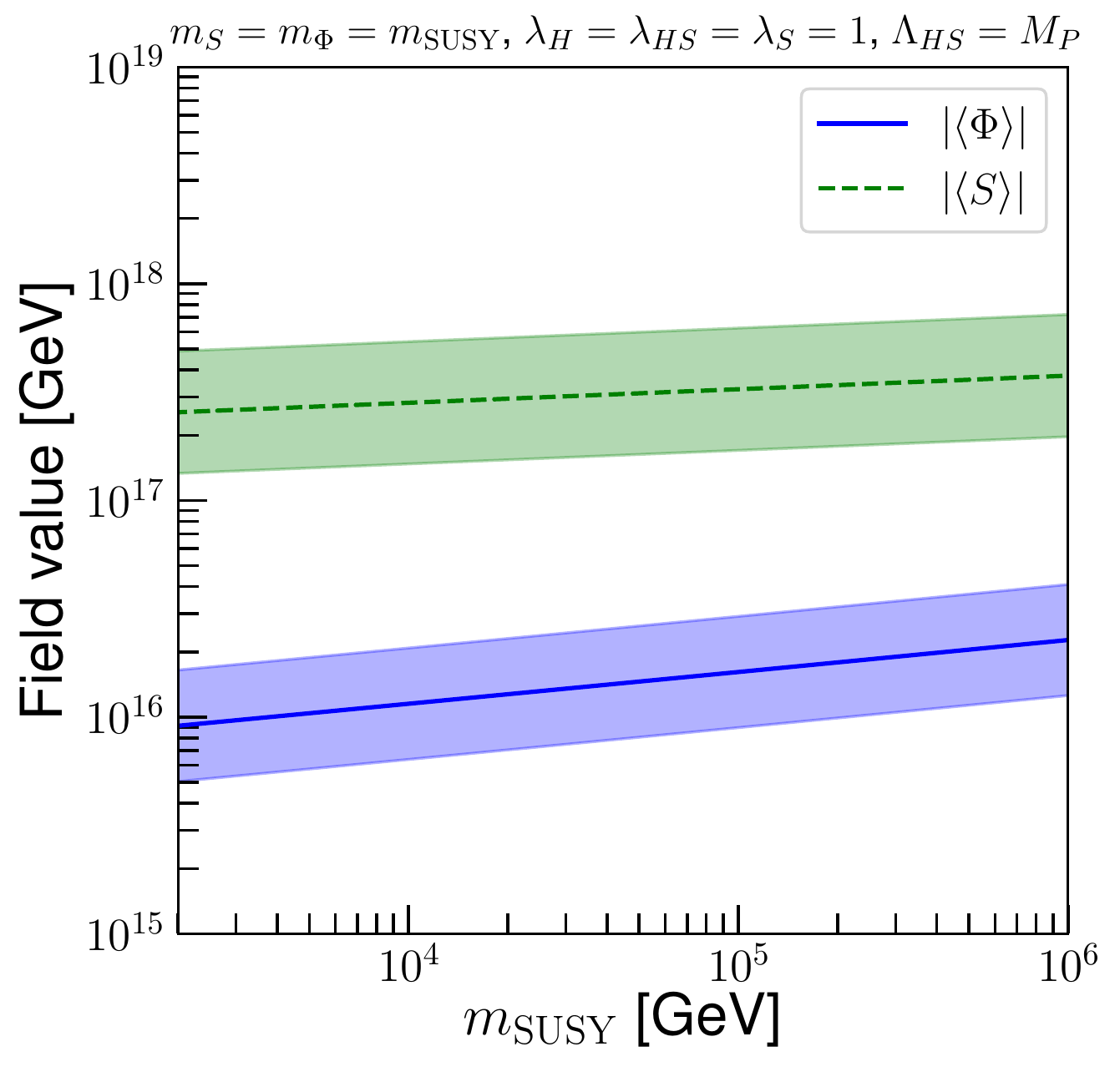}} 
  \caption{
    The absolute values of the VEVs $\langle S \rangle$ and $\langle \Phi \rangle$ as functions of $m_{\rm SUSY} = m_S  = m_{\Phi}$ in the blue solid and green dashed lines, respectively, where we take $\lambda_H = \lambda_{HS} = \lambda_S = 1$ and $\Lambda_{HS} = M_P$. Each band shows the range of the predicted values when we vary the parameter $\Lambda_{HS}$ from $M_P/2$ to $2 M_P$.
  }
  \label{fig:vev}
  \end{figure}

In Fig.~\ref{fig:vev}, we show the absolute values of the VEVs $\langle S \rangle$ and $\langle \Phi \rangle$ at the minimum of the potential~\eqref{eq:vofphiands} as functions of $m_{\rm SUSY} = m_S  = m_{\Phi}$ in the blue solid and green dashed lines, respectively, where we take $\lambda_H = \lambda_{HS} = \lambda_S = 1$ and $\Lambda_{HS} = M_P$. Each band shows the range of the predicted values when we vary the parameter $\Lambda_{HS}$ from $M_P/2$ to $2 M_P$. We see that $|\langle \Phi \rangle| \simeq 10^{16}$~GeV and $|\langle S \rangle|\simeq \text{a few} \times 10^{17}$~GeV in the range of $m_{\rm SUSY}$ shown in the plot. The dependence of these VEVs on the soft masses is mild since the terms in $V_F$ are composed of a large power of the fields $\Phi$ and $S$.

Notice that if there were operators of $H$, $\bar{H}$, and $S$ that have mass dimensions lower than those in Eq.~\eqref{eq:whs}, such as $(H\bar{H})^2$, the VEVs $|\langle \Phi \rangle|$ and $|\langle S \rangle|$ would be much smaller than those in Fig.~\ref{fig:vev} and thus phenomenologically disfavored. In the ordinary flipped SU(5) models, such dangerous operators are just assumed to be absent, whilst in our model their absence is attributed to the U(1)$_R$ symmetry.

The VEV $\langle \Phi \rangle = \langle \nu^c_H\rangle = \langle \nu^c_{\bar{H}}\rangle$ breaks the ${\rm SU}(5) \times {\rm U}(1)$ gauge group into the SM gauge group. The number of broken gauge symmetries is $25 - (8+3+1) = 13$, and thus among the $10 \times 2$ components of $H$ and $\bar{H}$, 13 components are absorbed by the gauge bosons, with 12 gauge bosons acquiring a mass $M_{V_5}$ and the remaining one a mass $M_{V_1}$. These masses are computed as 
\begin{align}
  M_{V_5} &= g_5 |\langle \Phi \rangle| ~, \qquad
  M_{V_1} = \sqrt{\frac{5}{2}} \biggl(\frac{24}{25} g_5^2 + \frac{1}{25} g_X^2\biggr)^{\frac{1}{2}} | \langle \Phi \rangle |~.
\end{align}

The remaining 7 components of $H$ and $\bar{H}$ appear as physical states. One of them corresponds to the flaton field $\Phi$, which acquires an ${\cal O}(m_{\rm SUSY})$ mass at the minimum of the potential. The other six components are $d_H^c$ and $d_{\bar{H}}^c$ in $H$ and $\bar{H}$, respectively. After $\nu^c_H$ and $\nu^c_{\bar{H}}$ develop a VEV, $d_H^c$ and $d_{\bar{H}}^c$ form vector-like mass terms with $D$ and $\bar{D}$ in $h$ and $\bar{h}$ via the first and second terms in Eq.~\eqref{eq:wdt}, with the masses 
\begin{align}
  M_{H_C} &=\lambda_4 |\langle \Phi \rangle| \biggl(\frac{\langle S \rangle}{\Lambda_{\rm DT}}\biggr)^8  \nonumber\\[2pt]
  &\simeq 
  \lambda_4 \times 7 \times 10^{11}  \times
  \biggl(\frac{|\langle \Phi \rangle|}{10^{16}~{\rm GeV}} \biggr) \biggl(\frac{\langle S \rangle}{3 \times 10^{17}~{\rm GeV}} \biggr)^8 
  \biggl(\frac{\Lambda_{\rm DT}}{10^{18}~{\rm GeV}}\biggr)^{-8}
  ~{\rm GeV}~,
  \\[2pt]
  M_{\bar{H}_C} &=\lambda_5 |\langle \Phi \rangle |
  ~,
\end{align}
respectively. As we see, $M_{H_C} \ll M_{\bar{H}_C}$ in our model; this can be contrasted with the case in the standard flipped SU(5), where $M_{H_C} \simeq M_{\bar{H}_C}$. The implications of the presence of a light color-triplet Higgs multiplet will be discussed below. 

Notice that even after the ${\rm SU}(5) \times {\rm U}(1)$ gauge group is broken, the MSSM doublets in $h$ and $\bar{h}$ do not acquire GUT-scale masses and only the color-triplet components become massive. This shows that the doublet-triplet splitting problem is solved in this setup by the missing partner mechanism~\cite{Masiero:1982fe, Grinstein:1982um, Antoniadis:1987dx}. We also note that for this mechanism to work, the absence of large $\mu$-term for $h$ and $\bar{h}$ is crucial. In the current setup, the U(1)$_R$ symmetry prohibits not only the bilinear operator $h \bar{h}$, but also all of the operators of the form $S^m (H\bar{H})^n h \bar{h}$, and thus there is no danger for the generation of a large $\mu$-term. This is another advantage in our model.

The VEVs of the $\Phi$ and $S$ fields also break the U(1)$_R$ symmetry. Since this is a global symmetry, a Nambu-Goldstone boson, called $R$-axion~\cite{Nelson:1993nf}, appears at low energies, which is a linear combination of the pseudo-scalar components of the U(1)$_R$-charged fields that develop VEVs. If this U(1)$_R$ symmetry is an exact symmetry, this $R$-axion is massless. In many realistic SUSY-breaking scenarios, however, the U(1)$_R$ symmetry is explicitly violated by a constant term in the superpotential, which is used to suppress the cosmological constant. We also assume an U(1)$_R$-breaking constant term in the superpotential, $W_{\rm const}$, in our model. Let us write the superpotential as $W = W_{0} + W_{\rm const}$, where $W_{0}$ denotes the superpotential that describes both the visible and hidden sectors. In the presence of $W_{\rm const}$, the $R$-axion acquires a mass through the supergravity effect. At the leading order in the $M_P^{-1}$ expansion, it is computed as~\cite{Bagger:1994hh}
\begin{equation}
  m_a^2 \simeq \frac{8}{f_a^2 M_P^2} \left|
  \left\langle W_{\rm const} \left[ K^i (K^{-1})_i^{~j} W^*_{0 j} - 3 W_0^* \right]\right\rangle 
  \right|~,
  \label{eq:ma2}
\end{equation} 
where $f_a$ is the decay constant of the $R$-axion, which is given by
\begin{equation}
  f_a = \left[ 2 K_i^{~j} r_ir_j \langle \phi_i \rangle \langle \phi_j \rangle \right]^{1/2} ~,
\end{equation}
with $r_i$ the $R$ charge of the field $\phi_i$, and 
\begin{equation}
  K^i \equiv \frac{\partial K}{\partial \phi_i} ~, 
  \quad 
  K_i^{~j} \equiv \frac{\partial^2 K}{\partial \phi^{*i} \phi_j} ~, 
  \quad 
  W_{0j}^* \equiv \frac{\partial W_0^*}{\partial \phi^{*j}} ~.
\end{equation}
Suppose that the VEVs of the U(1)$_R$ charged fields in the hidden sector are much smaller than $\langle S \rangle$. Then, we have 
\begin{align}
  f_a \simeq \frac{\sqrt{2}}{9} \langle S \rangle 
  = 
  5 \times 10^{16} \times 
  \biggl(\frac{\langle S \rangle}{3 \times 10^{17}~{\rm GeV}} \biggr) ~{\rm GeV}~.
\end{align}
To estimate the size of the right-hand side in Eq.~\eqref{eq:ma2}, we first note that on the assumption that the soft mass scale $m_{\rm SUSY}$ is generated via the gravity mediation effect from the hidden sector, the size of the $F$-term in the hidden sector, $F_Z$, is approximately given by $F_Z \sim m_{\rm SUSY} M_P$, and thus the typical energy scale of the hidden sector physics is $\Lambda_{\rm hid} \simeq \sqrt{F_Z} \simeq \sqrt{m_{\rm SUSY} M_P}$. As a result, the hidden-sector contribution to the terms in the square brackets in Eq.~\eqref{eq:ma2} is order 
\begin{equation}
  \Lambda_{\rm hid}^3 = \left( m_{\rm SUSY}  M_P\right)^{3/2} 
  \simeq 4 \times 10^{33} \times 
  \biggl(\frac{m_{\rm SUSY}}{10^{4}~{\rm GeV}}\biggr)^{\frac{3}{2}}
  ~{\rm GeV}^3 ~.
  \label{eq:lamz3}
\end{equation}
On the other hand, it is found that the contribution from $W_{HS}$ can well be approximated by that of the last term in Eq.~\eqref{eq:whs} and is given by 
\begin{equation}
  \left| \left\langle 
  S  \frac{\partial W^*_{HS}}{\partial S^*} - 3 W_{HS}^*\right\rangle \right|
  \simeq 5 \times 10^{38} \times  \biggl(\frac{\langle S \rangle}{3 \times 10^{17}~{\rm GeV}} \biggr)^{18} ~{\rm GeV}^3 ~,
\end{equation}
for $\Lambda_{HS} = M_P$,
which is much larger than that in Eq.~\eqref{eq:lamz3}.\footnote{We, however, find that $|F_S|, |F_H|, |F_{\bar{H}}| \ll |F_Z|$ due to the large VEVs of $S$, $H$, and $\bar{H}$, and thus the SUSY is broken mainly in the hidden sector. } Finally, $W_{\rm const}$ is estimated as 
\begin{align}
  W_{\rm const} \simeq |\langle W \rangle | = m_{3/2} M_P^2 
  = 6 \times 10^{40} \times \biggl(\frac{m_{3/2}}{10^{4}~{\rm GeV}}\biggr)~{\rm GeV}^3 ~,
\end{align}
where $m_{3/2}$ is the gravitino mass.
All in all, the $R$-axion mass $m_a$ is evaluated as 
\begin{align}
  m_a \simeq 1 \times 10^{5} \times \biggl(\frac{\langle S \rangle}{3 \times 10^{17}~{\rm GeV}} \biggr)^{8}\biggl(\frac{m_{3/2}}{10^{4}~{\rm GeV}}\biggr)^{1/2}~{\rm GeV} ~.
\end{align}
We see that $R$-axion is as heavy as other SUSY particles in this model. 

As we see above, the bilinear terms $H\bar{H}$ and $h\bar{h}$ are forbidden by the U(1)$_R$ symmetry in this model. The former is effectively induced from $W_{HS}$ after $S$, $H$, and $\bar{H}$ develop VEVs, with its coefficient, $\mu_H$, being ${\cal O}(m_{\rm SUSY})$. The generation of the $\mu$-term for $h\bar{h}$ depends on the U(1)$_R$ charges of the hidden sector fields. With the superpotential terms for the hidden sector having U(1)$_R$ charge $+2$, we expect there are U(1)$_R$ charged fields in the hidden sector. If the SUSY-breaking field $Z$ has U(1)$_R$ charge $19/9$, for example, the $\mu$ term is generated from a K\"{a}hler potential term of the form 
\begin{equation}
   \frac{1}{\Lambda}\int d^4 \theta \, Z^* h \bar{h} +{\rm h.c.} ~,
\end{equation}
with which we have $\mu = F_Z/\Lambda_Z \simeq {\cal O}(m_{\rm SUSY})$ for $\Lambda_Z \simeq M_P$. In general, a K\"{a}hler potential term comprised of $Z^*$, $S^{(*)}$, $(H\bar{H})^{(*)}$, and $h\bar{h}$ can contribute to the $\mu$-term if the operator is allowed by the U(1)$_R$ symmetry. Another contribution can be provided by the coupling to the superconformal compensator field $\Sigma$~\cite{Cremmer:1978hn, Cremmer:1982en, Kugo:1982mr, Kugo:1982cu}, which has Weyl weight $+1$ and U(1)$_R$ charge $+2/3$. With this field, there is a term allowed by the symmetry,
\begin{equation}
  \int d^4 \theta \, \frac{\Sigma^*}{\Sigma^2} S^* h\bar{h} + {\rm h.c.}~,
\end{equation}
which gives $\mu = {\cal O}(m_{3/2} \langle S \rangle/M_P)$. The $\mu$ term can also be generated at loop level; for example, one loop diagrams with Higgs bosons and a bino/wino yield a $\mu$ term that is suppressed by a loop factor compared with the bino/wino mass~\cite{Hall:2011jd}.  

Finally, we discuss the masses of right-handed neutrinos. Since right-handed neutrinos are charged under the $\text{SU}(5) \times \text{U}(1)$ and U(1)$_R$ symmetries, their masses can be generated only after these symmetries are broken. The dominant contribution to the masses comes from the superpotential term $W_{\rm neutrino}$ in Eq.~\eqref{eq:wneutrino}, which gives rise to a mass matrix for right-handed neutrinos of the form 
\begin{align}
  (M_R)_{ij} &= c_{ij} \frac{\langle S \rangle |\langle \Phi \rangle|^2}{\Lambda_{N}^2} \nonumber \\ &\simeq c_{ij} \times 3 \times 10^{13} \times 
  \biggl(\frac{\langle S \rangle}{3 \times 10^{17}~{\rm GeV}}\biggr)
  \biggl(\frac{|\langle \Phi \rangle|}{10^{16}~{\rm GeV}}\biggr)^2 
  \biggl(\frac{\Lambda_N}{10^{18}~{\rm GeV}}\biggr)^{-2}
  ~{\rm GeV}~.
\end{align}
For the first two generations, this size of the right-handed neutrino masses is large enough to explain the light neutrino masses via the seesaw mechanism~\cite{Minkowski:1977sc, Yanagida:1979as, Glashow:1979nm, GellMann:1980vs, Mohapatra:1979ia}, but may be too small for the third generation, as we see in Sec.~\ref{sec:flavor}. We, however, note that a relatively small value of $\Lambda_N$ may be sufficient to make the right-handed neutrino masses large enough. Moreover, the same operators as in Eq.~\eqref{eq:wneutrino} with a slightly lower effective cut-off scale can easily be generated by additional fields around the Planck scale; for instance, suppose that there are two singlet chiral superfields, {$\phi$ and $\bar{\phi}$}, 
with U(1)$_R$ charges $19/18$ and $17/18$, respectively, which have the superpotential interactions 
\begin{equation}
  W_{\rm ex} = \lambda_6^i F_i \bar{H} \phi + \mu_\phi \phi \bar{\phi} + \frac{1}{2} \kappa_\phi S \bar{\phi} \bar{\phi} ~. 
  \label{eq:wex}
\end{equation}
Below the mass scale of $\mu_\phi$, which is taken to be slightly below the Planck scale, these interactions generate an effective operator 
\begin{equation}
  W_{\rm neutrino}^\prime = \frac{\lambda_6^i \lambda_6^j \kappa_\phi}{2\mu_\phi^2} S(F_i^{\alpha\beta}\bar{H}_{\alpha\beta} )(F_j^{\gamma\delta}\bar{H}_{\gamma\delta})~,
  \label{eq:wneutrinop}
\end{equation}
which can generate a mass of ${\cal O} (10^{14})$~GeV for the third generation right-handed neutrino if $\mu_\phi/M_P = {\cal O}(0.1)$.

\subsection{Gauge coupling unification}

The gauge coupling constants of the ${\rm SU}(5) \times {\rm U}(1)$ interactions, $g_5$ and $g_X$, are matched onto the SM gauge couplings at the unification scale, $M_{\rm GUT} \simeq \langle \Phi \rangle$. At tree level, the matching conditions are 
\begin{align}
  g_2 (M_{\rm GUT}) = g_3 (M_{\rm GUT}) = g_5 (M_{\rm GUT}) ~,
  \label{eq:g23match}
\end{align}
and 
\begin{equation}
  \frac{25}{g_1^2 (M_{\rm GUT})} = \frac{1}{g_5^2 (M_{\rm GUT})}
  + \frac{24}{g_X^2 (M_{\rm GUT})} ~.
\end{equation}
We see that $g_1$ is not necessarily equal to $g_2$ and $g_3$ at the unification scale, contrary to the standard SU(5). We, however, note that if we require $g_5 (M_{\rm GUT}) = g_X (M_{\rm GUT})$, which is expected if the ${\rm SU}(5) \times {\rm U}(1)$ gauge group is embedded into a simple group such as SO(10),\footnote{We have chosen the normalization of the U(1) charge such that the equation $g_5 = g_X$ holds in this case. } then we need $g_1 (M_{\rm GUT}) = g_2 (M_{\rm GUT}) = g_3 (M_{\rm GUT})$ as in the standard SU(5). 

Note that in the present scenario $M_{\rm GUT}$ is different from that obtained in the MSSM, $\simeq 2 \times 10^{16}$~GeV, due to the existence of a light color-triplet Higgs multiplet at an intermediate scale. To see its effect on $M_{\rm GUT}$, let us investigate its contribution to the gauge coupling beta functions. The one-loop beta-function coefficients for the gauge couplings in the MSSM with and without the color-triplet Higgs multiplet, $b_a^{\text{MSSM} + H_C} $ and $b_a^{\text{MSSM}}$ ($a = 1,2,3$), respectively, are 
\begin{equation}
   b_a^{\text{MSSM} + H_C} = 
   \begin{pmatrix}
     7 \\ 1 \\ -2 
   \end{pmatrix}
   \quad \text{and} \quad 
   b_a^{\text{MSSM}} = 
   \begin{pmatrix}
     33/5 \\ 1 \\ -3 
   \end{pmatrix}
   ~.
   \label{eq:beta}
\end{equation}
Since the unification scale is determined by the condition~\eqref{eq:g23match}, the difference in $b_a^{\text{MSSM} + H_C} $ and $b_a^{\text{MSSM}}$ for $a = 2,3$ is relevant. As we see in Eq.~\eqref{eq:beta}, the light color-triplet Higgs multiplet increases the SU(3)$_C$ beta function, while it does not change the SU(2)$_L$ one. As a result, the unification point becomes higher than that in the MSSM, $\simeq 2 \times 10^{16}$~GeV, which may have some tension with the value of $|\langle \Phi \rangle|$ shown in Fig.~\ref{fig:vev}, as $M_{\rm GUT} > |\langle \Phi \rangle|$. The mismatch between these two scales may indicate that there are sizable threshold corrections at the scale $M_{\rm GUT}$ and thus the theory above this scale has a more complicated structure than our model, which is the case if the model is a part of the more fundamental theory based on a larger simple gauge group. A concrete model building for such a theory will be performed on another occasion~\cite{HHN}.

\subsection{$R$-parity violation}
\label{sec:rparityviolation}

The ${\rm SU}(5) \times {\rm U}(1)$ gauge symmetry by itself allows the renormalizable operators
\begin{align}
  W_{R\text{-odd}} = \mu^i_{FH} F_i^{\alpha\beta} \bar{H}_{\alpha\beta} 
  + \kappa^i_F\, \epsilon_{\alpha\beta\gamma\delta\epsilon} F_i^{\alpha\beta} H^{\gamma\delta} h^{\epsilon} 
  + \kappa^i_{\bar{f}} \,  \bar{f}_{i \alpha} H^{\alpha\beta} \bar{h}_{\beta} ~, 
\end{align}
which yield $R$-parity violating operators at low energies. All of these operators are eliminated by the U(1)$_R$ symmetry in our model. At non-renormalizable level, there are operators allowed by the symmetry; for instance, we have 
\begin{align}
  W_{R\text{-odd}} &= 
  \frac{c^i_{FH}}{\Lambda^6} (H\bar{H}) S^5 F_i^{\alpha\beta} \bar{H}_{\alpha\beta} 
  + \frac{c^i_F}{2\Lambda^4}\, \epsilon_{\alpha\beta\gamma\delta\epsilon} S^4 F_i^{\alpha\beta} H^{\gamma\delta} h^{\epsilon} 
  + \frac{\sqrt{2}c^i_{\bar{f}}}{\Lambda^4} \,  S^4\bar{f}_{i \alpha} H^{\alpha\beta} \bar{h}_{\beta} \nonumber \\[2pt]
  &+ \frac{\sqrt{2}c^{ijk}_{1}}{4\Lambda^6}  \epsilon_{\alpha \beta\gamma\delta\epsilon} S^5 F^{\alpha\beta}_i F^{\gamma\delta}_j H^{\epsilon \tau} \bar{f}_{k\tau}
  + \frac{c^{ijk}_{2}}{\sqrt{2}\Lambda^6}  \epsilon_{\alpha \beta\gamma\delta\epsilon} S^5 F^{\alpha\beta}_i H^{\gamma\delta} F^{\epsilon \tau}_j \bar{f}_{k\tau}
  \nonumber \\[2pt]
  & 
   + \frac{\sqrt{2} c^{ijk}_3}{\Lambda^6} S^5 \bar{f}_{i\alpha} \bar{f}_{j\beta} \ell^c_k H^{\alpha\beta} 
  + \dots ~,
  \label{eq:wodd}
\end{align}
where $\Lambda$ is a cut-off scale and the dots denote operators whose contribution is less significant. At low energies, these operators lead to 
\begin{align}
  W_{R\text{-odd}}^{\rm (eff)} &= 
  \frac{c^i_{FH}\langle \Phi\rangle^3 \langle S\rangle^5 }{\Lambda^6} \nu^c_i 
  + \frac{c^i_F \langle \Phi \rangle \langle S\rangle^4 }{\Lambda^4}\,  d^c_i\, D
  + \frac{c^i_{\bar{f}} \langle \Phi \rangle \langle S\rangle^4 }{\Lambda^4} \,  L_i \cdot H_u \nonumber \\[2pt]
  &- \frac{(c^{ijk}_{1}-c^{ijk}_{2}) \langle \Phi \rangle \langle S \rangle^5 }{\Lambda^6} \, 
   d_i^c\, Q_j \cdot L_k -\frac{c^{ijk}_{2} \langle \Phi \rangle \langle S\rangle^5 }{\Lambda^6}\, \epsilon^{abc} d_{ia}^c d_{jb}^c u^c_{kc} 
   \nonumber \\[2pt]
  &
   +\frac{c^{ijk}_{3} \langle \Phi \rangle \langle S\rangle^5 }{\Lambda^6}\, L_i \cdot L_j \, e_k^c +\dots ~,
\end{align}
where $a,b,c$ denote the color index, $\epsilon^{abc}$ is the totally antisymmetric tensor, and $A \cdot B \equiv \epsilon_{mn} A^m B^n$ with $m,n$ the SU(2)$_L$ index. Although these operators are suppressed by a large power of the cut-off scale $\Lambda$, some of them are actually incompatible with the limits on baryon/lepton-number violation~\cite{Allanach:2003eb, Chemtob:2004xr, Barbier:2004ez}. Among others, the bound on proton lifetime gives the most stringent constraint; for $\lambda_{ijk}^\prime L_i Q_j d_k^c$ and $(\lambda^{\prime \prime}_{ijk}/2) u^c_i d^c_j d^c_k$, we have, for instance~\cite{Barbier:2004ez} 
\begin{equation}
  |\lambda^\prime_{l1k}\lambda^{\prime \prime *}_{11k}| \lesssim 
 10^{-23} \times \biggl(\frac{m_{\tilde{d}_{k}^c}}{10~{\rm TeV}}\biggr)^2  ~,
\end{equation}
which strongly disfavors the operators in the second line in Eq.~\eqref{eq:wodd}. The bound from the washout of baryon asymmetry also restricts the $R$-parity violating operators~\cite{Campbell:1990fa, Fischler:1990gn, Dreiner:1992vm, Endo:2009cv, Dudas:2019gkj}. For example, we have~\cite{Endo:2009cv} 
\begin{equation}
  \frac{c^i_{\bar{f}} \langle \Phi \rangle \langle S\rangle^4 }{\Lambda^4}  \lesssim  {\cal O}(10^{-6}) \times |\mu| ~,
\end{equation}
which excludes the possibility that $c^i_{\bar{f}} = {\cal O}(1)$ even for $\Lambda = M_P$. 

Considering these severe limits, we introduce $R$-parity as an extra symmetry, under which the fields in our model transform as\footnote{When we consider the singlet fields in Eq.~\eqref{eq:wex}, we assign $R$-parity odd to $\phi$ and $\bar{\phi}$. }
\begin{equation}
  \begin{cases}
    \text{even}: & H, \, \bar{H}, \, h, \, \bar{h}, \, S
    \\[3pt]
    \text{odd}: & F_i, \, \bar{f}_i, \, \ell^c_i \,
  \end{cases}
  ~.
\end{equation}
With this assignment, all of the operators in Eq.~\eqref{eq:wodd} are prohibited, while the operators in Eqs.~(\ref{eq:wyukawa}--\ref{eq:whs}) are still allowed.

\section{Flavor structure}
\label{sec:flavor}

In this section, we examine the structure of the Yukawa terms in Eq.~\eqref{eq:wyukawa} and the neutrino mass matrix, following the discussion in Ref.~\cite{Ellis:1993ks}. We define the effective mass matrix for the field $F_i^{45}$ below the symmetry breaking scale as 
\begin{equation}
  W_R = M_R^{ij} F_i^{45} F_j^{45} ~.
  \label{eq:wr}
\end{equation}
As discussed in Sec.~\ref{sec:massspectrum}, we assume that it is generated by $W_{\rm neutrino}$ in Eq.~\eqref{eq:wneutrino} or some additional effect around the Planck scale, and that the values of $M_R^{ij}$ can be as large as ${\cal O}(10^{14})$~GeV.

We adopt the basis where $\lambda_2^{ij}$ is real and diagonal, without loss of generality: $\lambda_{2}^{ij} = \lambda_{2,D}^i \delta^{ij}$. In this basis, $\lambda^{ij}_1$ is a generic complex symmetric matrix and can be diagonalized with a unitary matrix, which we denote by $U_1$:
\begin{equation}
  \lambda_1 = U_1^T \lambda_{1,D} U_1^{} ~,
\end{equation}
where $\lambda_{1,D}$ is a diagonal matrix whose components are real and positive. Now it is convenient to separate out the phase factors in the matrix $U_1$ as many as possible, as done in Ref.~\cite{Ellis:1979hy}: $(U_1)_{ij} = e^{i\phi_{ij}} v_{ij}$, with $\phi_{ij}$ and $v_{ij}$ both real (summation over the indices $i,j$ is not taken). We then define the following diagonal phase matrices:
\begin{align}
  P_1 & = 
   {\rm diag} \left(e^{i \phi_{11}}, e^{i \phi_{21}}, e^{i \phi_{31}}  \right) ~,   \\
   P_2 &=  {\rm diag} \left(1, e^{i (\phi_{12} -\phi_{11})}, e^{i (\phi_{13} - \phi_{11})}  \right) ~.
\end{align}
By means of these matrices, we decompose the matrix $U_1$ as 
\begin{equation}
  U_1 = P_1 V^{\dagger} P_2 ~,
\end{equation}
to eliminate the phases in the first row and column in $U_1$. 
We further define $P \equiv P^2_1 \exp(-i 2\sum_i \phi_{i1}/3)$, for which ${\rm det}P = 1$. The phase factors $P_2$ and $\exp(-i \sum_i \phi_{i1}/3)$ can be absorbed by the field redefinition without changing the $\lambda_2$ couplings:
\begin{equation}
  F \to \exp(-i \sum_i \phi_{i1}/3) P_2^* F~, \qquad 
  \bar{f} \to \exp(i \sum_i \phi_{i1}/3)P_2\bar{f} ~.
\end{equation}
Since $U_1$ is a $3 \times 3$ unitary matrix, it has nine free parameters. $P_1$ and $P_2$ have five phases in total, and thus $V$ has four free parameters. With the condition ${\rm det} P = 1$, the phase matrix $P$ has two degrees of freedom.

In the above basis, the $\lambda_1$ matrix is written as 
\begin{equation}
  \lambda_1 = V^* P \lambda_{1,D} V^{\dagger} ~.
\end{equation}
After the field redefinition, the coupling matrix $\lambda_3$ has been transformed as
\begin{equation}
  \lambda_3 \to  \exp(i \sum_i \phi_{i1}/3) P_2 \lambda_3 ~.
\end{equation} 
We absorb the factor $\exp(i \sum_i \phi_{i1}/3) $ with the field redefinition 
\begin{equation}
  \ell^c \to  \exp(-i \sum_i \phi_{i1}/3) \ell^c ~.
\end{equation}
The remaining part, $P_2\lambda_3$, is diagonalized with unitary matrices $U_\ell$ and $U_{\ell^c}$ as 
\begin{equation}
 \lambda_{3,D}   = U_\ell^\dagger P_2 \lambda_3 U_{\ell^c}^{} ~.
\end{equation}
Finally, the matrix $M_R^{ij}$ in Eq.~\eqref{eq:wr} has been transformed as 
\begin{equation}
  M_R  \to \exp(-2i \sum_i \phi_{i1}/3) P_2^* M_R P_2^* \equiv M_R^\prime ~,
\end{equation}
which is diagonalized as 
\begin{equation}
  M_{R, D} = U_{\nu^c}^T M_R^\prime U_{\nu^c} ~,
\end{equation}
where $M_{R, D} \equiv {\rm diag}(M_{R_1}, M_{R_2}, M_{R_3})$ with $M_{R_i}$ real and positive. 

After these field redefinitions, the terms in $W_{\rm Yukawa}$ and $W_R$ are written as 
\begin{align}
  W_{\rm eff} &=\frac{1}{4} (V^* P\lambda_{1,D}V^\dagger )^{ij}\epsilon_{\alpha\beta\gamma\delta\epsilon} F_i^{\alpha \beta} F_j^{\gamma \delta} h^{\epsilon}_{} + \sqrt{2} \lambda_{2,D}^{i} F_i^{\alpha \beta}\bar{f}_{i\alpha}\bar{h}_{\beta} 
  \nonumber \\[2pt]
  &+ (U_\ell\lambda_{3,D} U_{\ell^c}^\dagger )^{ij}\bar{f}_{i\alpha}\ell^c_j h^{\alpha}
  + (U_{\nu^c}^* M_{R, D} U_{\nu^c}^\dagger )^{ij} F_i^{45} F_j^{45} ~.
\end{align}
Now we express the fields in the above equation in terms of the component fields defined as follows:
\begin{align}
  F^{ab}_i &= \frac{1}{\sqrt{2}} \epsilon^{abc} V_{ij} P^*_{jj} d^c_{jc} ~, \quad
  F^{an}_i = \frac{1}{\sqrt{2}} Q^{an}_i ~, \quad
  F^{45}_i =  \frac{1}{\sqrt{2}}  U^{ij}_{\nu^c} \nu^c_j ~, \nonumber \\[2pt]
  \bar{f}_{ia} &= u^c_{ia} ~, \quad \bar{f}_{in} = \epsilon_{nm} U_\ell^{*ij}  L_{j m} ~, 
  \quad \ell^c_i = U_{\ell^c}^{ij} e_j^c
\end{align}
where $m,n = 4,5$ denote the SU(2)$_L$ index, and 
\begin{align}
  h &= 
  \begin{pmatrix}
    D^1 \\ D^2 \\ D^3 \\ H_d^0 \\ H_d^- 
  \end{pmatrix} 
  ~, \quad
  \bar{h} = 
  \begin{pmatrix}
    \bar{D}_1 \\ \bar{D}_2 \\ \bar{D}_3 \\ H_u^0 \\ - H_u^+ 
  \end{pmatrix}
  ~.
\end{align}
We then have 
\begin{align}
  W_{\rm eff} &= 
  (\lambda_{1,D} V^\dagger)^{ij} d_i^c (Q_j \cdot H_d) 
  + \lambda^i_{2,D} u_i^c (Q_i \cdot H_u) 
  + \lambda_{3,D}^i (H_d \cdot L_i) e^c_i 
  \nonumber \\[2pt]
  & + (U^T_{\nu^c}\lambda_{2,D} U^*_\ell)^{ij} \nu_i^c (L_j \cdot H_u)  + \frac{1}{2} M_{R_i} \nu_i^c \nu^c_i 
  \nonumber \\[2pt]
  &- \frac{1}{2} (V^* P\lambda_{1,D}V^\dagger )^{ij} \epsilon_{abc} (Q^a_i \cdot Q^b_j) D^c 
  + (\lambda_{1,D}V^\dagger U_{\nu^c} )^{ij} d_i^c \nu_j^c D 
 \nonumber  \\[2pt]
 &+ (P^* V^T \lambda_{2,D} )^{ij} \epsilon^{abc} d_{ia}^c u_{jb}^c \bar{D}_c + (U_\ell^\dagger \lambda_{2,D} )^{ij} (L_i \cdot Q_j) \bar{D} 
 + (U_\ell \lambda_{3,D} )^{ij} u_i^c e_j^c D ~.
 \label{eq:weffcomp}
\end{align}
From the first two terms, we see that the up- and down-type quark masses are diagonalized if we define the components of the left-handed quark fields as 
\begin{equation}
  Q_i = \begin{pmatrix}
    u_i \\ V_{ij} d_j
  \end{pmatrix}
  ~.
\end{equation}
Therefore, the matrix $V$ is equal to the Cabibbo--Kobayashi--Maskawa (CKM) matrix. The eigenvalues of $\lambda_{1,D}$, $\lambda_{2,D}$, and $\lambda_{3,D}$ correspond to the down-type quark, up-type quark, and charged lepton Yukawa couplings, respectively. 

We here note in passing that the down-type quark and charged-lepton Yukawa couplings do not necessarily unify in flipped SU(5) GUT models, contrary to the standard SU(5). This feature could be advantageous given that the unification of Yukawa couplings is generically imperfect in the MSSM, especially for the first two generations. 

At the mass scale of right-handed neutrinos, we integrate out these fields to obtain the dimension-five operator for the light-neutrino mass matrix~\cite{Minkowski:1977sc, Yanagida:1979as, Glashow:1979nm, GellMann:1980vs, Mohapatra:1979ia}: 
\begin{align}
  W_\nu = 
  - \frac{1}{2} \left( U_\ell^\dagger \lambda_{2,D} U_{\nu^c} M_{R, D}^{-1} 
  U^T_{\nu^c} \lambda_{2,D} U_\ell^* \right)^{ij} (L_i \cdot H_u) (L_j \cdot H_u) ~.
\end{align}
For convenience, we define the neutrino mass matrix as 
\begin{equation}
  m_{\nu}^{ij} \equiv - \left(\lambda_{2,D} U_{\nu^c} M_{R, D}^{-1} 
  U^T_{\nu^c} \lambda_{2,D} \right)^{ij} \langle H_u^0 \rangle^2 ~. 
  \label{eq:seesaw}
\end{equation}
This mass matrix can be diagonalized with a unitary matrix: 
\begin{equation}
  m_\nu^D = U^T_\nu m_\nu U_\nu^{} ~.
\end{equation}
We express the left-handed lepton fields in terms of the mass eigenstates as 
\begin{equation}
  L_i = 
  \begin{pmatrix}
    U_{ij} \nu_j \\ e_i
  \end{pmatrix}
  ~,
\end{equation}
where $U$ is the Pontecorvo--Maki--Nakagawa--Sakata (PMNS) matrix. We then find that it is related to the unitary matrices $U_\ell$ and $U_\nu$ as 
\begin{equation}
  U = U_\ell^T U_\nu^{} ~. 
\end{equation}

Now recall that the couplings $\lambda_{2,D}^i$ correspond to the up-type quark Yukawa couplings. The eigenvalues of the mass matrix $m_\nu$ are, therefore, approximately given by 
\begin{equation}
  m_{\nu_i} \simeq \frac{m_{q_i}^2}{M_N} ~, 
  \label{eq:mnui}
\end{equation}
where $m_{q_i}$ are quark masses and $M_{N}$ represents the scale of right-handed neutrino masses. By requiring $m_{\nu_3} \lesssim 0.1$~eV, we obtain a rough lower limit on the right-handed neutrino mass scale: $M_N \gtrsim 3 \times 10^{14}$~GeV. The relation~\eqref{eq:seesaw} also suggests that the components in $m_{\nu}$ are hierarchical unless $M_{R,D}$ is hierarchical, and thus the matrix $U_\nu$ is close to the unity. In this case, the matrix $U_\ell$ can well be approximated by the PMNS matrix as 
\begin{equation}
  U_\ell \simeq U^T ~.
\end{equation}
We focus on this case in the following discussion.

\section{Proton decay}
\label{sec:protondecay}

In our model, proton decay is caused by the exchange of the color-triplet Higgs multiplets or the SU(5) gauge bosons; the former induces both the dimension-five and dimension-six operators, while the latter generates only the dimension-six operators. 

\subsection{Dimension-five proton decay}
\label{sec:dim5protondecay}

Let us first discuss the dimension-five proton decay operators induced by the color-triplet Higgs exchange. In many SUSY GUT models, such as the minimal SUSY SU(5)~\cite{Dimopoulos:1981zb, Sakai:1981gr}, these operators give the dominant contribution~\cite{Weinberg:1981wj, Sakai:1981pk}, with the main decay mode being $p \to K^+ \bar{\nu}$~\cite{Ellis:1981tv}. In the standard flipped SU(5), however, this contribution is extremely small since the $\mu$-term for $H\bar{H}$, $\mu_H$, is ${\cal O}(m_{\rm SUSY})$ and thus the coefficients of the dimension-five operators are suppressed by a factor of ${\cal O} (m_{\rm SUSY}/M_{\rm GUT}^2)$ due to the chirality flip. In our model, the suppression of the dimension-five proton decay by a small $\mu_H$ also exists, whereas one of the color-triplet Higgs multiplets has a mass much smaller than $M_{\rm GUT}$. It is, therefore, worth checking if the dimension-five proton decay rate is still small enough to evade the current experimental limits.

The exchange of the color-triplet Higgs multiplets is induced by the interactions in Eq.~\eqref{eq:weffcomp}. By integrating out the fields $D$, $\bar{D}$, $d^c_{H}$, and $d^c_{\bar{H}}$, we obtain the following dimension-five superpotential-type operators that violate baryon and lepton numbers:
\begin{align}
  W_5^{\rm eff} = 
  \frac{\mu_H}{M_{H_C} M_{\bar{H}_C}} &\biggl[
  \frac{1}{2} (V^* P \lambda_{1,D} V^\dagger)^{ij} (\lambda_{2,D} U_\ell^*)^{kl} \epsilon_{abc}  (Q_i^a \cdot Q_j^b) (Q_k^c \cdot L_l) 
  \nonumber \\
  &- (\lambda_{2,D} V P^*)^{ij} (U_\ell \lambda_{3,D})^{kl} 
  \epsilon^{abc} u^c_{ia} d^c_{jb} u^c_{kc} e^c_l 
  \biggr]~.
  \label{eq:w5eff}
\end{align}
As we see, these operators receive a suppression factor $\mu_H/M_{H_C}$ in addition to the ordinary factor $M^{-1}_{\bar{H}_C} \simeq M_{\rm GUT}^{-1}$ for the dimension-five operators induced at the GUT scale. To see if this suppression factor is sufficient to evade the current experimental bound, we note that in the minimal SUSY SU(5), the lifetime of $p \to K^+ \bar{\nu}$ is predicted to be $\tau (p \to K^+ \bar{\nu}) = {\cal O}(10^{30-31})$~years for the TeV-scale SUSY particles, the color-triplet Higgs mass of $\simeq 10^{16}$~GeV, and $\tan \beta = {\cal O}(1)$~\cite{Goto:1998qg, Murayama:2001ur}. Given that for $\mu_H \simeq 1$~TeV and $M_{H_C} \simeq 10^{11}$~GeV the suppression factor is $\mu_H/M_{H_C} = 10^{-8}$, we find that the proton lifetime in our model is much longer than the current experimental bound, $\tau (p \to K^+ \bar{\nu}) > 6.6 \times 10^{33}$~years~\cite{Abe:2014mwa, Takhistov:2016eqm}, for the same parameter choice.\footnote{Strictly speaking, this estimate is not accurate since the coefficients of the dimension-five operators in the minimal SUSY SU(5) have different flavor structure compared with those in Eq.~\eqref{eq:w5eff}. We, however, find that the coefficients are not so much different in these two cases, and thus our conclusion is unchanged. } If $m_{\rm SUSY}$ is much larger than the TeV scale, the suppression factor $\mu_H/M_{H_C}$ can be moderate. In this case, however, the dimension-five proton decay evades the experimental bound even in the minimal SUSY SU(5)~\cite{McKeen:2013dma, Liu:2013ula, Hisano:2013exa, Nagata:2013sba, Nagata:2013ive, Evans:2015bxa, Bajc:2015ita, Ellis:2015rya, Ellis:2016tjc, Ellis:2017djk, Evans:2019oyw, Ellis:2019fwf}, and thus there is no constraint on the the operators in Eq.~\eqref{eq:w5eff}. 

A potentially dangerous contribution to the dimension-five proton decay in our model is provided by the cut-off suppressed operators. We find that the following operators are allowed by the symmetries of the theory:
\begin{align}
  W_5^\prime &= \frac{c^{ijkl}_1}{\Lambda^2}  \epsilon_{\alpha\beta\gamma\delta\epsilon} S F_i^{\alpha\beta} F_j^{\gamma\delta} F_k^{\epsilon \tau} \bar{f}_{l \tau} 
  +\frac{c^{ijkl}_2}{\Lambda^2} S F_i^{\alpha\beta} \bar{f}_{j\alpha} \bar{f}_{k\beta} \ell^c_l ~.
  \label{eq:w5p}
\end{align}
It turns out that if $c_1^{ijkl}$ or $c_2^{ijkl}$ is ${\cal O}(1)$, then the rate of the dimension-five proton decay is too large for $m_{\rm SUSY} \lesssim {\cal O}(10^{8}) $~GeV, even for $\Lambda = M_P$~\cite{Dine:2013nga}. We, however, note that it is quite likely that the coefficients $c_1^{ijkl}$ and $c_2^{ijkl}$ are suppressed by the same mechanism as that explains the hierarchical structure of the SM quark and lepton Yukawa couplings. Indeed, if we consider a flavor-dependent U(1)$_R$ charge assignment to account for the Yukawa structure, then the coefficients $c_1^{ijkl}$ and $c_2^{ijkl}$ can be highly suppressed for $i, j, k, l = 1, 2$, with which we can evade the proton decay bound. In this paper, we just assume that a certain mechanism makes these coefficients small enough to evade the experimental limit, as often done in the literature.

\subsection{Dimension-six proton decay}
\label{sec:dim6protondecay}

The suppression of the dimension-five proton decay operators~\eqref{eq:w5eff} by a small $\mu_H$ is due to the chirality flip that is required to generate the operators. The exchange of the light color-triplet Higgs multiplet induces dimension-six operators as well via chirality-conserving diagrams. Since these dimension-six operators do not suffer from the suppression by $\mu_H$, they may dominate the contribution from $W_5^{\rm eff}$ even if they have larger mass dimensions than those in Eq.~\eqref{eq:w5eff}. In addition, the exchange of the SU(5) gauge bosons also induces the dimension-six proton decay operators, which gives the dominant contribution to proton decay in the standard flipped SU(5) model~\cite{Ellis:1988tx, Ellis:2002vk, Dorsner:2004xx, Li:2010dp, Ellis:2020qad}. We now study the dimension-six proton decay in our model, taking account of both the light color-triplet Higgs multiplet and SU(5) gauge boson exchange processes. 

The dimension-six proton decay operators are expressed as
\begin{equation}
  {\cal L}_{6}^{\rm eff}
 =C_{6(1)}^{i j k l}{\cal O}^{6(1)}_{i j k l}
 +C_{6(2)}^{i j k l}{\cal O}^{6(2)}_{i j k l}
 ~,
 \label{eq:l6eff}
 \end{equation}
 where 
 \begin{align}
  {\cal O}^{6(1)}_{i j k l}&=\int d^2\theta d^2\bar{\theta}~
 \epsilon_{abc}\epsilon_{mn}
 \bigl(u^{c\dagger}_i\bigr)^a
 \bigl(d^{c\dagger} _j\bigr)^b
 e^{-\frac{2}{3}g^\prime B}
 \bigl(e^{2g_3G}Q_k^m\bigr)^cL^n_l~,
  \\
 {\cal O}^{6(2)}_{i j k l}&=\int d^2\theta d^2\bar{\theta}
 \epsilon_{abc}\epsilon_{mn}~
 Q^{am}_iQ^{bn}_j
 e^{\frac{2}{3}g^\prime B}
 \bigl(e^{-2g_3G}u^{c\dagger} _k\bigr)^c
 e^{c\dagger} _l~,
 \end{align}
 with $G$ and $B$ the SU(3)$_C$ and U(1)$_Y$ gauge vector superfields,
 respectively.

 First, at the scale $M_{V_5}$, we integrate out the SU(5) gauge bosons to obtain these proton decay operators. The relevant gauge interaction terms are 
 \begin{align}
  K_{\rm gauge} =
 \sqrt{2}g_5\bigl(
 -\epsilon_{mn}(u^c_a)^{\dagger}X^m_a U_{\ell}^* L^n
 +\epsilon^{abc}(Q^{am})^{\dagger} X^m_bVP^* {d}^c_c
 + \epsilon_{mn}(\nu^c)^\dagger U_{\nu^c}^\dagger X^m_aQ^{an}
 +{\rm h.c.}
 \bigr)~,
 \label{eq:gaugeintflipped}
 \end{align}
 where $X_a^m$ denote the SU(5) gauge vector superfields. The tree-level exchange of the SU(5) gauge bosons then yields 
 \begin{align}
  C^{i j k l}_{6(1)}  (M_{V_5}) &= \frac{g_5^2}{M_{V_5}^2} (U^*_{\ell})_{il} V_{kj}^*  e^{i\varphi_j} = \frac{1}{|\langle \Phi \rangle|^2} (U_{\ell}^*)_{il} V_{kj}^*  e^{i\varphi_j}  ~,
 \nonumber \\
  C^{i j k l}_{6(2)} (M_{V_5}) & = 0 ~,
 \label{eq:dim6gutmatchflipped}
 \end{align}
 where $\varphi_i$ are the phases in the phase matrix $P$, defined by $P_{ij} \equiv e^{i \varphi_i} \delta_{ij}$. The contribution of the exchange of the heavier color-triplet Higgs multiplet is much smaller than the one in Eq.~\eqref{eq:dim6gutmatchflipped} due to the suppression by small Yukawa couplings. We therefore neglect this contribution in the following analysis.

 These Wilson coefficients are evolved down to the mass scale of $M_{H_C}$ using the renormalization group equations (RGEs). The renormalization factor can be computed using the results in Ref.~\cite{Munoz:1986kq} at one-loop level\footnote{The two-loop RGEs for these coefficients are also available~\cite{Hisano:2013ege}.} as 
 \begin{equation}
  C^{i j k l}_{6(1)}  (M_{H_C}) = A_1^H C^{i j k l}_{6(1)}  (M_{V_5}) ~,
 \end{equation}
 with 
 \begin{equation}
  A_1^H =  \biggl[
    \frac{\alpha_3(M_{H_C})}{\alpha_3(M_{V_5})}
    \biggr]^{\frac{2}{3}}
    \biggl[
    \frac{\alpha_2(M_{H_C})}{\alpha_2(M_{V_5})}
    \biggr]^{-\frac{3}{2}}
    \biggl[
    \frac{\alpha_1(M_{H_C})}{\alpha_1(M_{V_5})}
    \biggr]^{-\frac{11}{210}} ~.
 \end{equation}
At this scale, we integrate out the light color-triplet Higgs at tree level, which contributes only to the coefficient $C^{ijkl}_{6(2)} $:
\begin{equation}
  C^{i j k l}_{6(2)} (M_{H_C}) = - \frac{1}{2 M_{H_C}^2} 
  (V^* P \lambda_{1,D} V^\dagger)^{ij} (U_{\ell}^* \lambda_{3,D})^{kl} ~.
  \label{eq:ccolorhiggscontr}
\end{equation}
We then run both of these coefficients down to the hadronic scale. The resultant renormalization factors are given as 
\begin{align}
  C^{i j k l}_{6(1)}  (\mu_{\rm had}) &= A_1^R C^{i j kl}_{6(1)}  (M_{H_C}) ~,  \nonumber \\[2pt]
  C^{i j k l}_{6(2)}  (\mu_{\rm had}) &= A_2^R C^{i jkl}_{6(2)}  (M_{H_C}) ~,  
\end{align}
with 
\begin{align}
  A_{1}^R &=
  \biggl[
 \frac{\alpha_3(m_{\text{SUSY}})}{\alpha_3(M_{H_C})}
 \biggr]^{\frac{4}{9}}
 \biggl[
 \frac{\alpha_2(m_{\text{SUSY}})}{\alpha_2(M_{H_C})}
 \biggr]^{-\frac{3}{2}}
 \biggl[
 \frac{\alpha_1(m_{\text{SUSY}})}{\alpha_1(M_{H_C})}
 \biggr]^{-\frac{1}{18}}  
 \nonumber \\
 &\times
 \biggl[
 \frac{\alpha_3(m_Z)}{\alpha_3(m_{\rm SUSY})}
 \biggr]^{\frac{2}{7}}
 \biggl[
 \frac{\alpha_2(m_Z)}{\alpha_2(m_{\rm SUSY})}
 \biggr]^{\frac{27}{38}}
 \biggl[
 \frac{\alpha_1(m_Z)}{\alpha_1(m_{\rm SUSY})}
 \biggr]^{-\frac{11}{82}} A_L ~, \nonumber \\
  A_{2}^R &=
  \biggl[
 \frac{\alpha_3(m_{\text{SUSY}})}{\alpha_3(M_{H_C})}
 \biggr]^{\frac{4}{9}}
 \biggl[
 \frac{\alpha_2(m_{\text{SUSY}})}{\alpha_2(M_{H_C})}
 \biggr]^{-\frac{3}{2}}
 \biggl[
 \frac{\alpha_1(m_{\text{SUSY}})}{\alpha_1(M_{H_C})}
 \biggr]^{-\frac{23}{198}}
 \nonumber \\
 &\times
 \biggl[
 \frac{\alpha_3(m_Z)}{\alpha_3(m_{\rm SUSY})}
 \biggr]^{\frac{2}{7}}
 \biggl[
 \frac{\alpha_2(m_Z)}{\alpha_2(m_{\rm SUSY})}
 \biggr]^{\frac{27}{38}}
 \biggl[
 \frac{\alpha_1(m_Z)}{\alpha_1(m_{\rm SUSY})}
 \biggr]^{-\frac{23}{82}} A_L ~,
 \end{align}
where $\mu_{\rm had}$ is the hadronic scale, for which we use $\mu_{\rm had} = 2$~GeV. The renormalization factors below the SUSY-breaking scale are computed from the results in Ref.~\cite{Abbott:1980zj}. $A_L$ is the long-distance QCD renormalization factor, which is computed in Ref.~\cite{Nihei:1994tx} at the two-loop level: $A_L = 1.24$. 

In the standard flipped SU(5), $p \to \pi^0 e^+/\mu^+$ are the main decay channels. The relevant effective operators for these modes are 
\begin{align}
 {\cal L}(p\to \pi^0 \ell^+_i)
&=C_{RL}(udu \ell_i)\bigl[\epsilon_{abc}(u_R^ad_R^b)(u_L^c\ell_{Li}^{})\bigr]
+C_{LR}(udu\ell_i)\bigl[\epsilon_{abc}(u_L^ad_L^b)(u_R^c\ell_{Ri}^{})\bigr]
~,
\label{eq:lagptopil}
\end{align}
where 
\begin{align}
 C_{RL}(udu\ell_i)&=C^{111i}_{6(1)}~, \nonumber \\
 C_{LR}(udu\ell_i)&=V_{j1}\bigl[
C^{1j1i}_{6(2)}+C^{j11i}_{6(2)}
\bigr]~.
\end{align}
The partial decay width can then be computed as 
\begin{equation}
 \Gamma (p\to  \pi^0 \ell^+_i)=
\frac{m_p}{32\pi}\biggl(1-\frac{m_\pi^2}{m_p^2}\biggr)^2
\bigl[
\vert {\cal A}_L(p\to \pi^0 \ell^+_i) \vert^2+
\vert {\cal A}_R(p\to \pi^0 \ell^+_i) \vert^2
\bigr]~,
\label{eq:aptopil}
\end{equation}
where we have neglected the lepton mass for simplicity,
$m_p$ and $m_\pi$ denote the masses of the proton and pion, respectively, and 
\begin{align}
 {\cal A}_L(p\to \pi^0 \ell^+_i)&=
C_{RL}(udu\ell_i)\langle \pi^0\vert (ud)_Ru_L\vert p\rangle_{\ell_i}
~,\nonumber \\
 {\cal A}_R(p\to \pi^0 \ell^+_i)&=
C_{LR}(udu\ell_i)\langle \pi^0\vert (ud)_Ru_L\vert p\rangle_{\ell_i}
~.
\end{align}
For the matrix elements $\langle \pi^0\vert (ud)_Ru_L\vert p\rangle_{\ell_i}$, we use the values obtained with QCD lattice simulations in Ref.~\cite{Aoki:2017puj}: 
\begin{align}
  {\langle \pi^0|(ud)_Ru_L|p\rangle_e} &= -0.131(4)(13) ~{\rm GeV}^2 ~,\\[2pt]
{\langle \pi^0|(ud)_Ru_L|p\rangle_\mu} &= -0.118(3)(12) ~{\rm GeV}^2~,
\end{align}
where the first and second parentheses show the statistical and systematic
uncertainties, respectively, and the subscript $\ell_i = e, \mu$ shows that the matrix element is computed for the final state with a charged antilepton $\ell_i^+$.

By using the above equations, we can express the partial decay widths of the $p \to \pi^0 e^+/\mu^+$ decay modes as 
\begin{align}
  \Gamma (p\to  \pi^0 e^+)&=
\frac{m_p}{32\pi}\biggl(1-\frac{m_\pi^2}{m_p^2}\biggr)^2
\left|V_{ud}\right|^2
\left|\left( U_\ell \right)_{11}\right|^2
\left( {\langle \pi^0|(ud)_Ru_L|p\rangle_e} \right)^2 
\nonumber \\[2pt]
& \times \biggl[
\frac{(A^H_1)^2 (A_1^R)^2}{|\langle \Phi \rangle|^4} + 
\frac{(A_2^R)^2}{M_{H_C}^4}
\left\{ y_d (M_{H_C}) y_e (M_{H_C}) \right\}^2 
\biggr] ~, \label{eq:gamppie} \\[3pt]
\Gamma (p\to  \pi^0 \mu^+)&=
\frac{m_p}{32\pi}\biggl(1-\frac{m_\pi^2}{m_p^2}\biggr)^2
\left|V_{ud}\right|^2
\left|\left( U_\ell \right)_{12}\right|^2
\left( {\langle \pi^0|(ud)_Ru_L|p\rangle_\mu} \right)^2 
\nonumber \\[2pt]
& \times \biggl[
\frac{(A^H_1)^2 (A_1^R)^2}{|\langle \Phi \rangle|^4} + 
\frac{(A_2^R)^2}{M_{H_C}^4}
\left\{ y_d (M_{H_C}) y_\mu (M_{H_C}) \right\}^2 
\biggr] ~,  \label{eq:gamppimu}
\end{align}
where $y_d$, $y_e$, and $y_\mu$ are the Yukawa couplings of down quark, electron, and muon, respectively. As we see, these decay widths do not depend on the phases $\varphi_i$ in $P$. In addition, the SU(5) gauge boson and color-triplet Higgs multiplet exchange processes do not interfere with each other since the chirality of the charged lepton in the final state is different and thus these two processes are in principle distinguishable.\footnote{Strictly speaking, for $p\to \pi^0\mu^+$, there is a correction which depends on the interference at ${\cal O}(m_\mu/m_p)$. However, its effect is negligible except for a narrow parameter region around $|{\cal A}_L|\simeq |{\cal A}_R|$, and is at most $2m_\mu/m_p\simeq 0.2$ even for $|{\cal A}_L| = |{\cal A}_R|$.
\label{ft:mumass}
}

Since the color-triplet Higgs exchange process is induced by the Yukawa interactions, its contribution tends to be more significant for the decay modes that contain the second generations. In addition, this process contributes to only the decay channels that contain a charged lepton. Motivated by this observation, we also consider the $p \to K^0 e^+ / \mu^+ $ decay modes. The effective interactions in this case are given by
\begin{align}
 {\cal L}(p\to K^0 \ell^+_i)
&=C_{RL}(usu\ell_i)\bigl[\epsilon_{abc}(u_R^as_R^b)(u_L^c\ell_{Li}^{})\bigr]
+C_{LR}(usu\ell_i)\bigl[\epsilon_{abc}(u_L^as_L^b)(u_R^c\ell_{Ri}^{})\bigr]
~,
\label{eq:lagptokl}
\end{align}
with
\begin{align}
 C_{RL}(usu\ell_i)&=C^{121i}_{6(1)}~, \nonumber \\
 C_{LR}(usu\ell_i)&=V_{j2}\bigl[C^{1j1i}_{6(2)}
+C^{j11i}_{6(2)}
\bigr]~.
\end{align}
We then obtain the partial decay width
\begin{equation}
 \Gamma (p\to  K^0 \ell^+_i)=
\frac{m_p}{32\pi}\biggl(1-\frac{m_K^2}{m_p^2}\biggr)^2
\bigl[
\vert {\cal A}_L(p\to K^0 \ell^+_i) \vert^2+
\vert {\cal A}_R(p\to K^0 \ell^+_i) \vert^2
\bigr]~,
\label{eq:gamptokefl}
\end{equation}
where $m_K$ is the kaon mass and 
\begin{align}
 {\cal A}_L(p\to K^0 \ell^+_i)&=
C_{RL}(usu\ell_i)\langle K^0\vert (us)_Ru_L\vert p\rangle_{\ell_i}
~,\nonumber \\
 {\cal A}_R(p\to K^0 \ell^+_i)&=
C_{LR}(usu\ell_i)\langle K^0\vert (us)_Ru_L\vert p\rangle_{\ell_i}
~.
\end{align}
The matrix elements are again computed in Ref.~\cite{Aoki:2017puj}:
\begin{align}
  {\langle K^0|(us)_Ru_L|p\rangle_e}   &=  0.103(3)(11)~{\rm GeV}^2~,  \\[2pt]
{\langle K^0|(us)_Ru_L|p\rangle_\mu}   &=  0.099(2)(10)~{\rm GeV}^2 ~.
\end{align}

By using these results, we obtain
\begin{align}
  \Gamma (p\to  K^0 e^+)&=
\frac{m_p}{32\pi}\biggl(1-\frac{m_K^2}{m_p^2}\biggr)^2
\left|V_{us}\right|^2
\left|\left( U_\ell \right)_{11}\right|^2
\left( {\langle K^0|(us)_Ru_L|p\rangle_e}  \right)^2 
\nonumber \\[2pt]
& \times \biggl[
\frac{(A^H_1)^2 (A_1^R)^2}{|\langle \Phi \rangle|^4} + 
\frac{(A_2^R)^2}{M_{H_C}^4}
\left\{ y_s (M_{H_C}) y_e (M_{H_C}) \right\}^2 
\biggr] ~, \label{eq:gampke} \\[3pt]
\Gamma (p\to  K^0 \mu^+)&=
\frac{m_p}{32\pi}\biggl(1-\frac{m_K^2}{m_p^2}\biggr)^2
\left|V_{us}\right|^2
\left|\left( U_\ell \right)_{12}\right|^2
\left( {\langle K^0|(us)_Ru_L|p\rangle_\mu} \right)^2 
\nonumber \\[2pt]
& \times \biggl[
\frac{(A^H_1)^2 (A_1^R)^2}{|\langle \Phi \rangle|^4} + 
\frac{(A_2^R)^2}{M_{H_C}^4}
\left\{ y_s (M_{H_C}) y_\mu (M_{H_C}) \right\}^2 
\biggr] ~, 
\label{eq:gampkmu}
\end{align}
where $y_s$ is the strange quark Yukawa coupling. Again, these decay widths are independent of the phases $\varphi_i$, and there is no interference between the SU(5) gauge boson and color-triplet Higgs contributions (see, however, footnote~\ref{ft:mumass} for an ${\cal O}(m_\mu /m_p)$ correction). 

As we see in Eqs.~\eqref{eq:gamppie}, \eqref{eq:gamppimu}, \eqref{eq:gampke}, and \eqref{eq:gampkmu}, the proton decay rates depend on the unitary matrix $U_\ell$. Motivated by the discussion in Sec.~\ref{sec:flavor}, we set $U_\ell = U^T$ with $U$ the PMNS matrix; \textit{i.e.}, $(U_\ell)_{11} = \cos \theta_{12} \cos \theta_{13}$ and $(U_\ell)_{12} = - \sin \theta_{12} \cos \theta_{23} - \cos \theta_{12} \sin \theta_{23} \sin \theta_{13} e^{i \delta}$, for the normal ordering, where $\theta_{12}$, $\theta_{23}$, $\theta_{13}$ are the neutrino mixing angles and $\delta$ is the neutrino Dirac CP phase. We use the values obtained by the NuFIT~5.0 global analysis of neutrino oscillation measurements for these parameters~\cite{Esteban:2020cvm, NuFIT5_0}, assuming the normal ordering. For other input parameters, we use the values recommended by the Particle Data Group~\cite{Tanabashi:2018oca}. 

\begin{figure}
  \centering
  \subcaptionbox{\label{fig:ptopie}
  $p \to \pi^0 e^+$
  }
  {\includegraphics[width=0.48\textwidth]{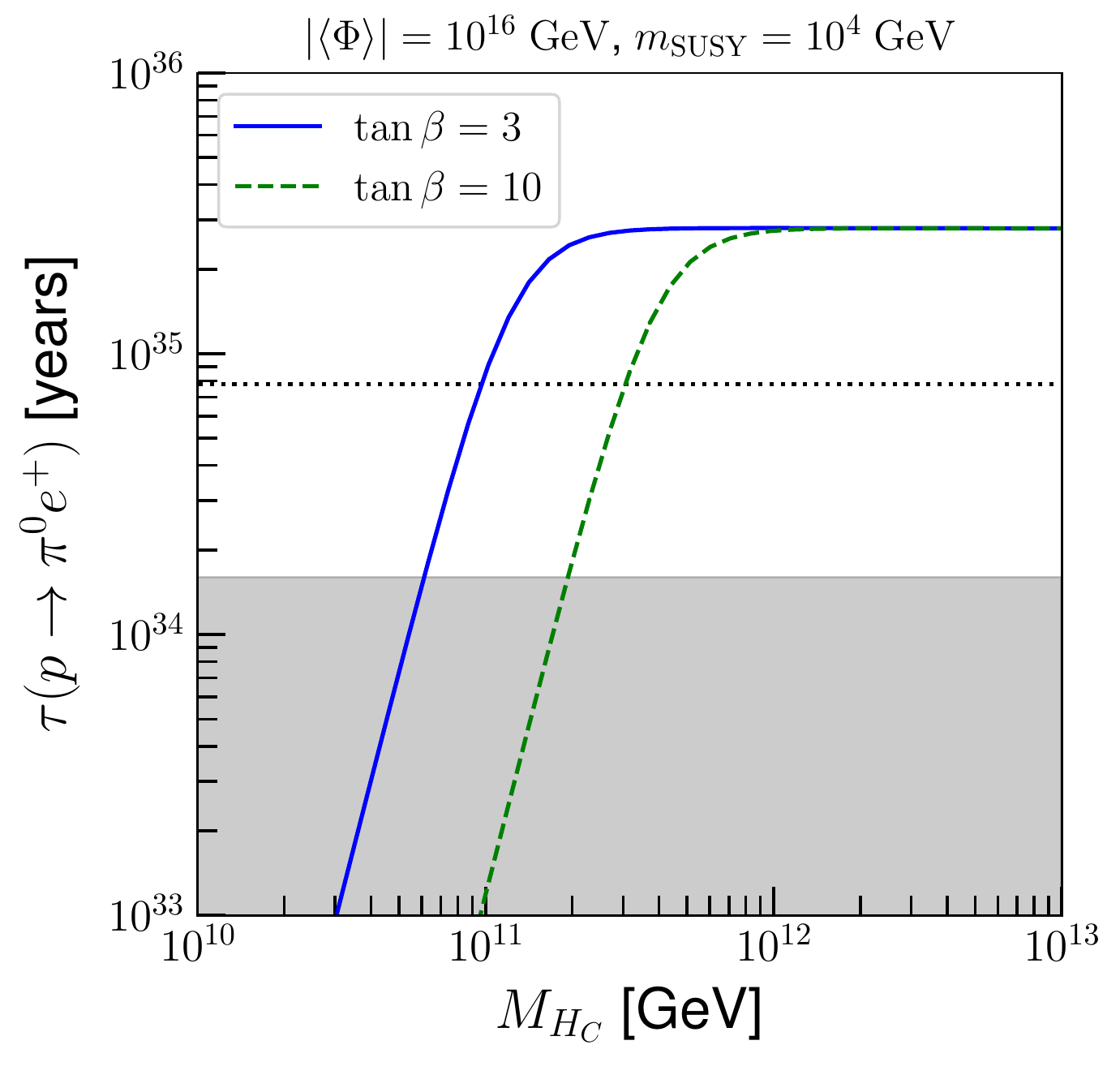}}
  \subcaptionbox{\label{fig:ptopimu}
  $p \to \pi^0 \mu^+$
    \vspace{10mm}
  }
  { 
  \includegraphics[width=0.48\textwidth]{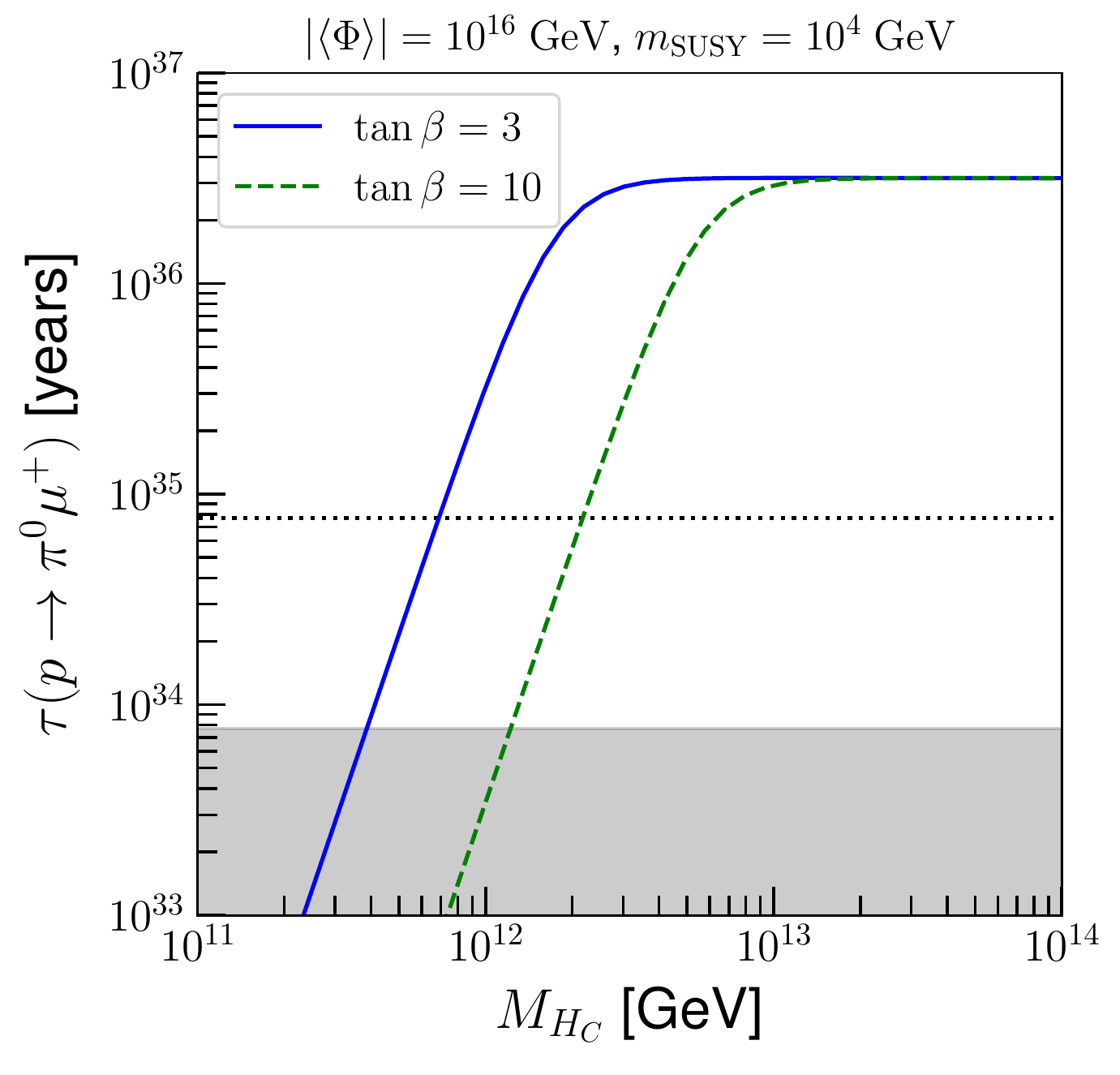}}
  \subcaptionbox{\label{fig:ptoke}
  $p \to K^0 e^+$
  }
  {\includegraphics[width=0.48\textwidth]{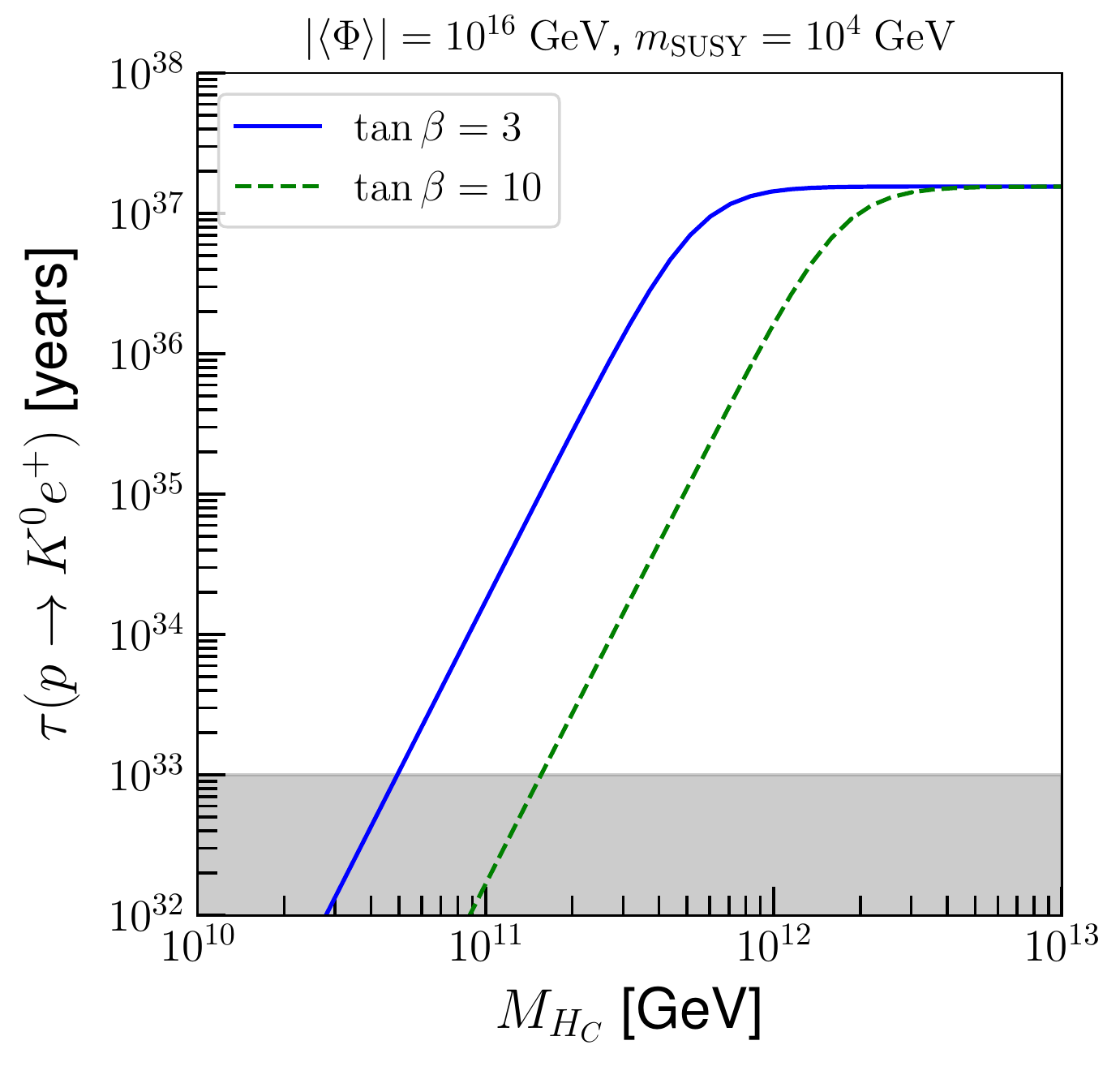}}
  \subcaptionbox{\label{fig:ptokmu}
  $p \to K^0 \mu^+$
  }
  { 
  \includegraphics[width=0.48\textwidth]{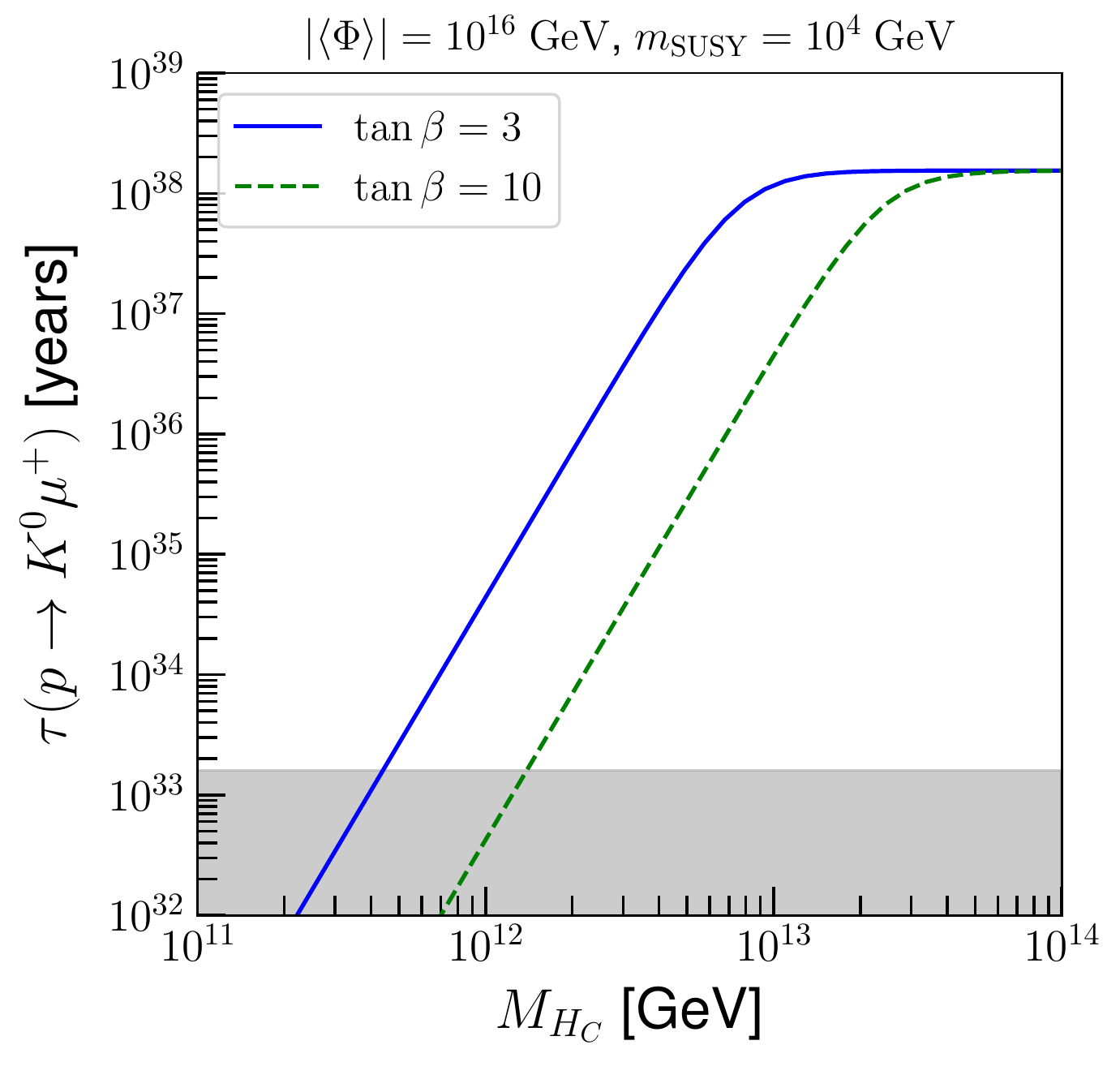}}
  \caption{
    The lifetime of each decay mode as a function of the color-triplet Higgs mass $M_{H_C}$ for $|\langle \Phi \rangle | = 10^{16}~{\rm GeV}$ and $m_{\rm SUSY} = 10^4$~GeV. The solid blue and dashed green lines correspond to $\tan \beta \equiv \langle H^0_u \rangle/ \langle H^0_d \rangle = 3$ and 10, respectively. The gray shades represent the current experimental limits from Super-Kamiokande~\cite{Miura:2016krn, Kobayashi:2005pe, Regis:2012sn}, and the horizontal black dotted lines show the sensitivities of Hyper-Kamiokande~\cite{Abe:2018uyc}. 
  }
  \label{fig:lifetime}
  \end{figure}

In Fig.~\ref{fig:lifetime}, we show the lifetime of each decay mode as a function of the color-triplet Higgs mass $M_{H_C}$ for $|\langle \Phi \rangle | = 10^{16}~{\rm GeV}$ and $m_{\rm SUSY} = 10^4$~GeV. The solid blue and dashed green lines correspond to $\tan \beta \equiv \langle H^0_u \rangle/ \langle H^0_d \rangle = 3$ and 10, respectively. The current experimental limits from Super-Kamiokande, $\tau (p \to \pi^0 e^+) > 1.6 \times 10^{34}$~years~\cite{Miura:2016krn}, $\tau (p \to \pi^0 \mu^+) > 7.7 \times 10^{33}$~years~\cite{Miura:2016krn}, $\tau (p \to K^0 e^+) > 1.0 \times 10^{33}$~years~\cite{Kobayashi:2005pe}, and $\tau (p \to K^0 \mu^+) > 1.6 \times 10^{33}$~years~\cite{Regis:2012sn}, are shown in the gray shaded area, and the expected 90\% CL limits from the 10-year run of Hyper-Kamiokande~\cite{Abe:2018uyc} are shown in the horizontal black dotted lines (if available). As we see, $p \to \pi^0 \mu^+$ and $p \to K^0 \mu^+$ are most sensitive to the proton decay induced by the light color-triplet Higgs exchange, setting a lower limit on its mass, $M_{H_C} \gtrsim 5 \times 10^{11}$~GeV for $\tan\beta\gtrsim 3$.
The lifetime is shorter for the decay mode that contains a second-generation quark or lepton in the final state due to their larger Yukawa couplings. It is also found that the decay rate increases for a larger $\tan \beta$, since both the down-type quark and charged lepton Yukawa couplings are inversely proportional to $\cos \beta$. For a sufficiently large $M_{H_C}$, the lifetimes approach asymptotic values, which are determined by the SU(5) gauge boson exchange process. 

We note that we can in principle distinguish the proton decay induced by the color-triplet Higgs exchange from that by the SU(5) gauge boson exchange in our model by measuring the chirality of the final-state lepton. As we see in Eqs.~\eqref{eq:dim6gutmatchflipped} and \eqref{eq:ccolorhiggscontr}, the SU(5) gauge boson generates only the operator that contains a left-handed lepton,\footnote{This is a specific feature of the flipped SU(5); in the case of the standard SU(5), for instance, the SU(5) gauge bosons contribute to the both types of the operators. } while the color-triplet Higgs exchange induces that with a right-handed lepton. We can thus discriminate these two contributions by investigating the polarization of the antilepton in the decay products, offering an interesting way to test this model.

Our model predicts that if the mass of the lighter color-triplet Higgs field is $\lesssim 10^{12}$~GeV, then proton decay can be discovered at Hyper-Kamiokande in the $p \to \pi^0 \mu^+$ channel, and presumably in the $p \to K^0 \mu^+$ channel as well. Notice that without the color-triplet Higgs exchange contribution, {\it i.e.}, in the case of the standard flipped SU(5), $\tau (p \to \pi^0 e^+) < \tau (p \to \pi^0 \mu^+) < \tau (p \to K^0 \mu^+) $ as can be seen from Fig.~\ref{fig:lifetime}. In the minimal SUSY SU(5), on the other hand, the most promising decay channel is either $p \to K^+ \bar{\nu}$ or $p \to \pi^0 e^+$, depending on the sfermion mass scale~\cite{Nagata:2013sba, Nagata:2013ive, Ellis:2019fwf}. It is worth noting that $p \to K^+ \bar{\nu}$ does not occur in the standard flipped SU(5)~\cite{Ellis:1993ks, Ellis:2020qad}, as well as in our model since the lighter color-triplet Higgs multiplet does not couple to the left-handed neutrinos. In any cases, the $p \to \pi^0 \mu^+$ and $p \to K^0 \mu^+$ modes are usually subdominant, and therefore offer a promising way to distinguish our model from others, which highly motivates to search for the $p \to \pi^0 \mu^+$ and $p \to K^0 \mu^+$ modes in future proton decay experiments, in addition to the standard channels $p \to \pi^0 e^+ $ and $p \to K^+ \bar{\nu}$.

\section{Conclusion and discussion}
\label{sec:conclusion}

We have constructed a flipped SU(5) SUSY GUT model which is invariant under a global U(1)$_R$ symmetry---except the constant term in the superpotential. This U(1)$_R$ symmetry eliminates Higgs potential terms up to sufficiently higher mass dimensions such that the SU(5)-breaking Higgs fields acquire a GUT-scale VEV in the flat direction after SUSY is spontaneously broken. In addition, the $\mu$-terms of the Higgs fields are also forbidden by the U(1)$_R$ symmetry and, in particular, the MSSM Higgs superfields naturally have a mass much smaller than the GUT scale through the missing partner mechanism. The masses of  right-handed neutrinos in our model are generated by non-renormalizable operators suppressed by a cut-off scale, with which we can obtain a light neutrino mass spectrum and mixing that are consistent with neutrino oscillation data.

In this model, one of the color-triplet Higgs multiplets lies around an intermediate scale, since its mass is generated by a higher-dimensional operator. As we have seen in Sec.~\ref{sec:dim6protondecay}, this mass, $M_{H_C}$, is stringently constrained by the current limit on proton decay lifetime provided by the Super-Kamiokande experiments; $M_{H_C} \gtrsim 5 \times 10^{11}$~GeV for $\tan\beta\gtrsim 3$. In particular, it is found that the $p \to \pi^0 \mu^+$ and $p \to K^0 \mu^+$ modes are most sensitive to the light color-triplet Higgs exchange process. Future proton decay experiments at Hyper-Kamiokande can probe $M_{H_C} \lesssim 10^{12}$~GeV in these decay channels. Since these decay channels tend to be subdominant in other GUT models, they offer a promising way to distinguish our model from, \textit{e.g.}, the standard/flipped SU(5) GUT models.

There are several interesting possibilities to extend our model to address other problems. For instance, we may consider a flavor-dependent U(1)$_R$ charge assignment to explain the structure of the Yukawa couplings. In fact, in our model, the ratio $\langle S \rangle / \Lambda$ is predicted to be ${\cal O}(0.1)$ for $\Lambda \simeq M_P$, which allows us to take $\langle S \rangle / \Lambda \simeq 0.22 \simeq |V_{us}|$, as often done in models that use the Froggatt--Nielsen mechanism~\cite{Froggatt:1978nt} to explain the Yukawa structure. Another direction is to embed our model into a theory based on a larger simple gauge group---\textit{e.g.}, SO(10)---to achieve a complete unification of gauge interactions. These more ambitious model buildings will be explored in future work~\cite{HHN}.

\section*{Acknowledgments}

This work is supported in part by the Grant-in-Aid for Innovative Areas (No.19H05810 [KH], No.19H05802 [KH], No.18H05542 [NN]), Scientific Research B (No.20H01897 [KH and NN]), and Young Scientists B (No.17K14270 [NN]). 

\section*{Appendix}
\appendix

\section{Model with a discrete $R$-symmetry}
\label{sec:discreter}

Instead of the U(1)$_R$ symmetry discussed in the text, we may consider a discrete $R$ symmetry to eliminate unwanted terms. In this Appendix, we show an example for such models.

We assume that the model considered here possesses a $\mathbb{Z}_{17}$ $R$ symmetry. The matter content of this model is the same as that of the U(1)$_R$ version, as summarized in Table~\ref{tab:modeldisc} with the charge assignment. We also assume this model to have an exact $R$-parity to suppress dangerous terms such as $F_i \bar{H}$. The superpotential terms allowed by the symmetries in this case are 
\begin{align}
  W &=\frac{1}{4} \lambda_1^{ij}\epsilon_{\alpha\beta\gamma\delta\epsilon} F_i^{\alpha \beta} F_j^{\gamma \delta} h^{\epsilon}_{} + \sqrt{2} \lambda_2^{ij} F_i^{\alpha \beta}\bar{f}_{j\alpha}\bar{h}_{\beta} + \lambda_3^{ij}\bar{f}_{i\alpha}\ell^c_j h^{\alpha} \nonumber \\[2pt]
  &+ \frac{\lambda_4}{4\Lambda_{\rm DT}^7} \epsilon_{\alpha\beta\gamma\delta\epsilon} S^7 H^{\alpha\beta}H^{\gamma\delta} h^{\epsilon} + \frac{1}{4} \lambda_5\epsilon^{\alpha\beta\gamma\delta\epsilon}  \bar{H}_{\alpha\beta}\bar{H}_{\gamma\delta} \bar{h}_{\epsilon} \nonumber
  \\[2pt]
 &+ \frac{c_{ij}}{2\Lambda_{N}^2} S (F_i^{\alpha\beta}\bar{H}_{\alpha\beta} )(F_j^{\gamma\delta}\bar{H}_{\gamma\delta}) 
 + \frac{\lambda_{H}}{4\Lambda_{HS}^{5}} (H^{\alpha\beta} \bar{H}_{\alpha\beta})^4 
 + \frac{\lambda_{HS}}{12\Lambda_{HS}^{7}} (H^{\alpha\beta} \bar{H}_{\alpha\beta})^3 S^4
 \nonumber
  \\[2pt]
 &
   + \frac{\lambda_{HS}^\prime}{16 \Lambda_{HS}^{9}} (H^{\alpha\beta} \bar{H}_{\alpha\beta})^2 S^8 
   + \frac{\lambda_{HS}^{\prime \prime}}{12 \Lambda_{HS}^{11}} (H^{\alpha\beta} \bar{H}_{\alpha\beta}) S^{12}
   + \frac{\lambda_S}{16 \Lambda_{HS}^{13}} S^{16} ~.
\end{align}

\begin{table}[t]
  \centering
  \begin{tabular}{l l cccc}
  \hline \hline
  Fields & Components & SU(5)\quad & U(1)\quad & $\mathbb{Z}_{17R}$ & $R$-parity \\
  \hline
  $F_i$ & $d^c_i$, $Q_i$, $\nu^c_i$ & $\mathbf{10}$ & $+1$ & $1$ & $-$ \\ 
  $\bar{f}_i$ & $u^c_i$, $L_i$ & $\overline{\mathbf{5}}$ & $-3$ &$1$& $-$\\ 
  ${\ell}_i^c$ & $e^c_i$ & ${\mathbf{1}}$ & $+5$ &$1$ & $-$\\ 
  $H$ & $d^c_H$, $Q_H$, $\nu^c_H$ & $\mathbf{10}$ & $+1$ &$8$ & $+$\\ 
  $\bar{H}$ & ${d}^c_{\bar{H}}$, ${Q}_{\bar{H}}$, ${\nu}^c_{\bar{H}}$ & $\overline{\mathbf{10}}$ & $-1$  &$1$ & $+$\\ 
  $h$ & $D$, $H_d$ & $\mathbf{5}$ & $-2$ &$0$ & $+$\\ 
  $\bar{h}$ & $\bar{D}$, $H_u$ & $\overline{\mathbf{5}}$ & $+2$ &$0$ & $+$\\ 
  $S$ & $S$ & ${\mathbf{1}}$ & $0$ &$-2$ & $+$\\ 
   \hline
  \hline
  \end{tabular}
  \caption{
   The field content and the charge assignments in the $\mathbb{Z}_{17}$ $R$ symmetric model. 
  }
  \label{tab:modeldisc}
  \end{table}

A characteristic feature of this type of models, compared with the U(1)$_R$ models, is that $R$-axion does not appear after the $R$ symmetry is spontaneously broken. In addition, generically speaking, there are more allowed terms than in the U(1)$_R$ models, which may be advantageous for some cases. For example, we have 
\begin{equation}
  W_\mu = \frac{\lambda_\mu }{16 \Lambda_{\mu}^{15}} S^{16} h\bar{h}~, 
\end{equation}
which can generate a $\mu$-term of ${\cal O}(m_{\rm SUSY})$.


\bibliographystyle{utphysmod}
\bibliography{ref}


\end{document}